# Electrochemical Strain Microscopy with Blocking Electrodes: The Role of Electromigration and Diffusion

A.N. Morozovska,[1] E.A. Eliseev,[2] and S.V. Kalinin[3,*]

[1] Institute of Semiconductor Physics, National Academy of Sciences of Ukraine,
41, pr. Nauki, 03028 Kiev, Ukraine

[2] Institute for Problems of Materials Science, National Academy of Sciences of Ukraine,
3, Krjijanovskogo, 03142 Kiev, Ukraine

[3] The Center for Nanophase Materials Sciences, Oak Ridge National Laboratory,
Oak Ridge, TN 37831

Electrochemical strains are a ubiquitous feature of solid state ionic devices ranging from ion batteries and fuel cells to electroresistive and memristive memories. Recently, we proposed a scanning probe microscopy (SPM) based approach, referred as electrochemical strain microscopy (ESM), for probing local ionic flows and electrochemical reactions in solids based on bias-strain coupling. In ESM, the sharp SPM tip concentrates the electric field in a small (10-50 nm) region of material, inducing interfacial electrochemical processes and ionic flows. The resultant electrochemical strains are determined from dynamic surface displacement and provide information on local electrochemical functionality. Here, we analyze image formation mechanism in ESM for a special case of mixed electronic-ionic conductor with blocking tip electrode, and determine frequency dependence of response, role of diffusion and electromigration effects, and resolution and detection limits.

[*] Sergei2@ornl.gov

## I. Introduction

Ionic transport and electrochemical processes in solids directly underpin a broad variety of energy conversion and storage technologies ranging from Li-ion[1] and Li-air[2] batteries to solid oxide fuel cells (SOFC).[3,4] Beyond energy applications, the applications of solid state ionic systems include electrochemical sensors and gas pumps,[5,6] as well as several classes of emerging information technology devices such as non-volatile electroresistive[7,8,9] and memristive[10] memories. Ionic phenomena are intrinsically linked to the operation of ferroelectric[11] and oxide electronic devices,[12] and may determine optimal synthetic strategies[13,14] and control e.g. failure and fatigue[15] mechanisms. Notably, many oxides extensively studied in the context of condensed matter physics of strongly correlated systems such as manganites, cobaltite, and ferrites,[16,17] are also broadly used in SOFC applications as electrodes, electrolytes, or bipolar plate materials. Given that characteristic diffusion lengths are typically much smaller, ionic phenomena can strongly affect and complement interpretations within framework of purely physical models.[18,19,20]

Advancements in solid state ionic materials and devices necessitate the understanding of mechanisms of ionic flows and electrochemical reactivity. Unlike the solution-based electrochemistry that offers the advantage of (lateral) spatial uniformity within bulk or surface layers and allows for use of ultramicroelectrodes, the solid materials are characterized by the presence of broad set of extended and localized defects such as dislocations, antiphase boundaries, morphological features, surface terminations. Correspondingly, understanding ionic processes in solids requires the capability for local probing, ideally at the length scale from ~0.5 - 1 nm level of individual structural defect to micron scale levels of mesoscopic device or signal generation of Raman or optical microscopy.

Recently, we have demonstrated the strong bias-strain coupling mediated by electrochemical processes in ionic materials can be used as a basis for high resolution imaging and spectroscopy of ionic and electrochemical phenomena in nanometer-scale volumes. Following the initial demonstration of electrochemical strain microscopy for layered Li-intercalation material,[21] this approach was further illustrated for mapping ionic dynamics and decoupling of reaction and transport phenomena in Si anode.[22] The theoretical limits of the electrochemical strain detection were compared to the classical technique based on Faradaic current detection in Ref. [23], demonstrating potential for $10^6$-$10^8$ decrease in probing volume.

Despite the progress in experimental studies, the fundamental mechanisms of electrochemical strain microscopy remain relatively unexplored. Here, we develop the analytical description of ESM mechanism for the ionically-blocking electrode for a special case of mixed ionic-electronic conductor (MIEC) with equal electron and mobile carrier concentration, determine the frequency dependence of response and roles of diffusion and electromigration effects, and establish resolution and sensitivity limits. This analysis can also be applied in a more general context of the strain induced by biased planar or disc electrodes on MIEC surfaces.[24,25]

## II. Mechanisms of Electrochemical Strain Microscopy

The application of electric bias to the SPM tip can induce interfacial electrochemical process on the tip-surface junction. Traditionally, these processes are explored through formation of new phases that can be identified by SPM topography imaging or e.g. Raman imaging [26,2728,29,30,31] and are generally irreversible.

Here, we aim to explore the dynamic effects associated with ***reversible*** electrochemical reaction-transport processes at the tip-surface junction, which generally precede the large-scale irreversible processes. In this case, electrochemical process at the tip-surface junction may result in change of concentration of mobile species, **Fig. 1 (a)**. These move through the material due to the concentration gradients (diffusion) and tip-induced electric field (electromigration). Note that the bias-induced redistribution of ionic species is possible even when the interfacial reaction is suppressed, i.e. the probe is completely blocking. The changes in ionic concentration in material results in electrochemical strains due to Vegard (concentration change, see e.g. Ref. [32]) and deformation potential (i.e. changes in oxidation state) effects.[33] These strains will result in the deformation of free surface, which are directly detectable through the deflection of scanning probe microscopy (SPM) tip. The detection limits of ~2-5 pm over ~5-50 nm lateral area at ~10-100 kHz frequencies can be achieved. Note that similar bias-strain detection principle, as well as detection limits, is realized in Piezoresponse Force Microscopy of ferroelectric and multiferroic materials, and recent advances in understanding of elementary mechanism of polarization switching and domain dynamics in these materials are available.[34,35,36]

The potential distribution in ESM tip-surface junction is illustrated in Fig. 1 (b). Here, the curve (I) corresponds to the general case where both interfacial reaction induced by potential drop at tip surface junction and electromigration contribute to the ESM signal. Note that potential drop at the tip-surface junction can be also due to dielectric gap effect;[37] however, in the absence of electrochemical process the problem is reduced to purely electrostatic one. Curve II corresponds to the purely diffusion-controlled mechanism, in which all potential drop occurs in tip-surface junction and electric field does not penetrate the material, similar to the case of supporting electrolyte in liquid-phase electrochemistry.[38,39] This situation is close to the extensively studied problem of diffusion-strain coupling in electroactive nanoparticles.[40,41, 42]

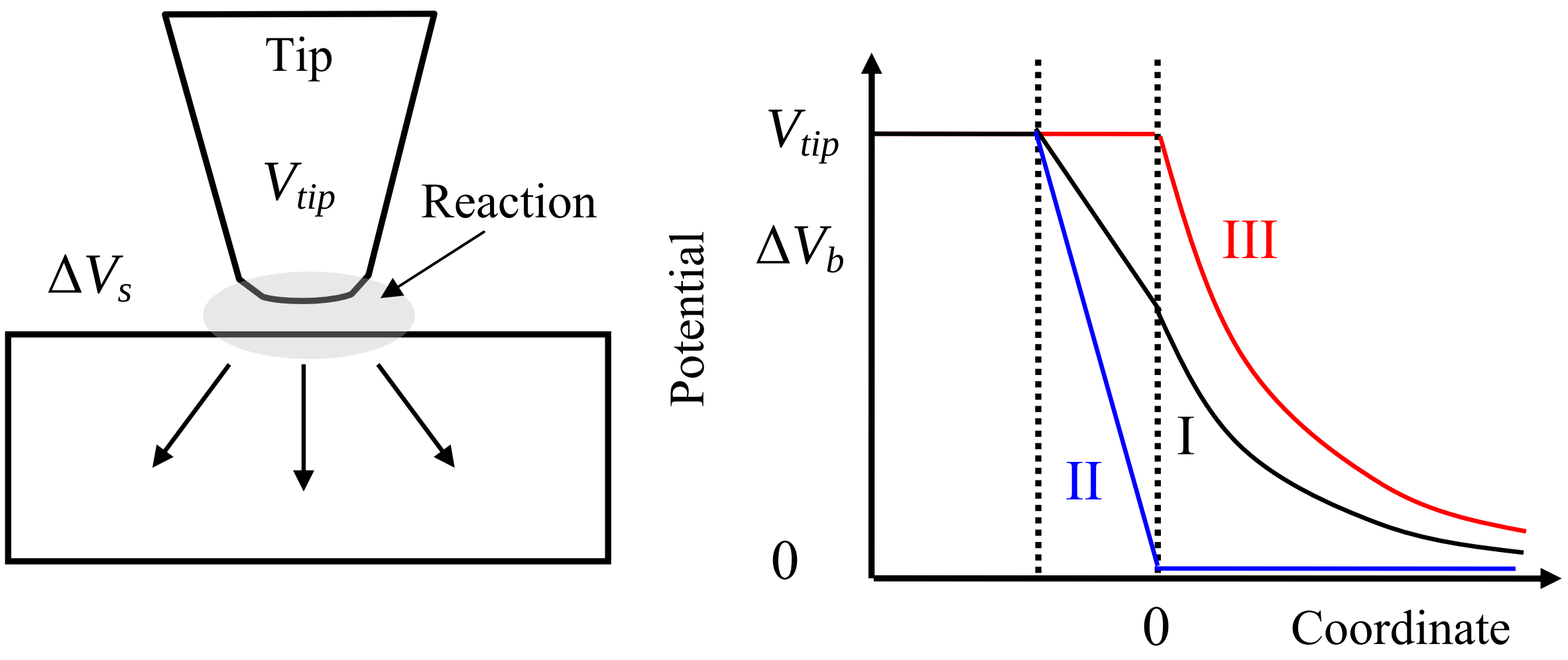

**Fig. 1.** **(a)** Schematics of the electrochemical processes at the tip-surface junction in Electrochemical Strain Microscopy. Application of the bias, $V_{tip}$, to the SPM probe leads to potential drops in the tip-surface junction an din the bulk, $V_{tip} = \Delta V_s + \Delta V_{b.}$ The potential drop in the junction can lead to the electrochemical reaction and generation of mobile ionic species that redistribute under the combined effect of electric field and concentration gradients. (b) Limiting cases for potential distribution under the tip. Note that the potential drop in the bulk will be non-linear even in the absence of ionic and electronic screening as a consequence of the localized nature of the probe.

The analytical solution for this diffusion-controlled case in Electrochemical Strain Microscopy was recently developed.[43] Finally, case III corresponds to the case of fully blocking electrode, where the ionic dynamics in solid is driven purely by electric field generated by the tip and the total amount of mobile species within material remains constant. This case is likely to be realized for low tip-potentials below the onset of interfacial electrochemical process (note that polarization at nanoscale electrodes can be much larger then at macroscopic ones due to smaller number of nucleation sites) and high frequencies, and is analyzed in details below.

## III. Mechanism of ESM with blocking probe

### III.1. Problem statement

To model the ESM signal, we consider the case of semi-infinite mixed ionic-electronic conductive (MIEC) material with mobile ionized donors and electrons. Corresponding concentration fields are $N_d^+(\mathbf{r})$ and $n(\mathbf{r})$, respectively. We consider the case when the extant donors in the material are neutral or singly ionized. The neutral donors are *immobile*, whereas the charged ones are *mobile*[44]. Correspondingly, in this case the total amount of mobile donors is equal to total amount of conductive electrodes, and the analysis of more complicated case of strong background conductance or ion concentration is deferred to future studies. The approximation of semi-infinite MIEC is valid when its thickness $h$ is much higher than the screening radius $R_D$. Typical values of $R_D$ for MIECs are ~0.5-50 nm, but it depends on the material parameters. Note, that MIEC thickness 10 times more than $R_D$ is quite enough for the semi-infinite approximation validity, since the ESM probe electric field decay exponentially with the depth z as $\exp(-z/R_D)$.

Due to the rotational isotropy of the tip-surface junction, we define the problem in the 2D cylindrical coordinates. The electric potential $\varphi(\mathbf{r})$ under the tip can be found self-consistently in the quasi-static approximation from the boundary-value problem:

$$\left(\frac{\partial^2}{\partial z^2}+\frac{1}{\rho}\frac{\partial}{\partial \rho}\left(\rho\frac{\partial}{\partial \rho}\right)\right)\varphi(\mathbf{r})=-\frac{q}{\varepsilon_0\varepsilon}\left(N_d^+(\mathbf{r})-n(\mathbf{r})\right), \tag{1a}$$

$$\varphi(\rho, z=0)=V_0(\rho,t), \quad \varphi(\rho, z\to\infty)=0\,, \tag{1b}$$

where $\mathbf{r}=\{x,y,z\}$ is the radius vector and $\rho=\sqrt{x^2+y^2}$ is the polar radius, $q$=1.6·10$^{-19}$C is the electron charge, $\varepsilon_0$ = 8.854 10$^{-12}$F/m is the dielectric permittivity of vacuum, ε is relative dielectric permittivity of the MIEC. The axially symmetric electric bias $V_0(\rho,t)$ represents the effect of the biased tip electrode.

To define the boundary condition representative of the ESM experiment, we assume that the tip potential $V_0(\rho,t)$ is constant inside the circle of radius $R_0$ and zero outside (*shielded-probe model*). This condition provides an approximate description of the probe tip having a well-defined characteristic size. We further utilize the fact approximate solutions developed here are insensitive to the details of the probe shape. Note, that the boundary condition $\varphi(\rho,z=0)=V_0(\rho,t)$ is also applicable for the case of the screening charge migration[45] or conductive water droplet[46,47] at the tip-surface junction. The difference between these cases is that for ideal tip surface contact that contact radius is typically 3-20 nm and can be determined from classical indentation theory, and strains are detected over full contact region. For water droplet case, the electrical contact area can be significantly higher (0.1 – 1 μm), and strains are detected only over the area of mechanical tip-surface contact.

For the **small periodic** electric potential $V_0(\rho,t)\sim V_0(\rho,\omega)\exp(i\omega t)$, we assume that the space charge concentration can be represented as

$$N_d^+(\mathbf{r},t)=\overline{N}_d^+ +\delta N_d^+(\mathbf{r},t), \qquad (2a)$$

$$n(\mathbf{r},t)=\overline{n}+\delta n(\mathbf{r},t), \qquad (2b)$$

Here $\overline{N}_d^+$ and $\overline{n}$ are the equilibrium concentrations for the semi-infinite MIEC, which can be found from the static metal-MIEC contact problem with relevant boundary conditions.

Here, we analyze a specific case of MIEC material in which electron and mobile donor concentrations are equal, i.e. defect equilibrium of the form $N_d^0 \leftrightarrow N_d^+ + e$ takes place. The case in which there is a large concentration of non-mobile donors or background electronic or ionic conductivity is present will be treated elsewhere, Ref. [45]. For the semi-infinite MIEC without Schottky barrier at $z=0$ (i.e. $\varphi(\rho,z=0)=0$, $\varphi(\rho,z\to\infty)=0$) the equilibrium concentrations are constant and equal: $\overline{N}_d^+=\overline{n}$, the static internal field is zero (see **Appendix A**). The solutions for Schottky profiles will be analyzed elsewhere (see e.g.

Ref.[45]). The periodic variations $\delta N_d^+(\mathbf{r},t)$ and $\delta n(\mathbf{r},t)$ are caused by the small periodic electric potential $\varphi(\mathbf{r},t)$ at the contact.

Generally, the analysis gives rise to the non-linear coupled equations for electric currents and electric potentials. Here, we use the linear drift-diffusion model for the ionic and electronic currents, $J_d \approx D_d \nabla N_d^+ - \eta_d N_d^+ \nabla\varphi$ and $J_n \approx D_n \nabla n - \eta_d n \nabla\varphi$. We assume constant diffusion coefficients $D_{d,n}$ and mobilities $\eta_{n,d}$, as discussed in Refs. [48, 49].

Within the model Eq.(2) kinetic Planck-Nernst-Einstein equations (see e.g. Refs.[50, 51]) acquire the form:

$$\Delta\varphi(\mathbf{r},t) = -\frac{q}{\varepsilon_0\varepsilon}\left(\delta N_d^+(\mathbf{r},t) - \delta n(\mathbf{r})\right), \tag{3a}$$

$$\frac{\partial N_d^+}{\partial t} + \frac{1}{q}\operatorname{div}\mathbf{J}_d = \frac{\partial \delta N_d^+}{\partial t} - D_d \Delta\delta N_d^+ - \eta_d \overline{N}_d^+ \Delta\varphi = 0, \tag{3b}$$

$$-\frac{\partial n}{\partial t} + \frac{1}{q}\operatorname{div}\mathbf{J}_n = -\frac{\partial \delta n}{\partial t} + D_n \Delta\delta n - \eta_n \overline{n}\Delta\varphi = 0. \tag{3c}$$

In Eqs. (3b,c) we omitted the terms $\nabla\delta N_d^+ \nabla\varphi$ and $\nabla\delta n \nabla\varphi$ as proportional to $V_0^2(\rho,\omega)\exp(2i\omega t)$, which correspond to the first order of perturbation theory valid for small voltages $\left|qV_0/k_B T\right| << 1$. Also, here we use the decoupling approximation and ignore the strain, deformation potential and flexoelectric effect[52] contributions in Eq. (3).

Note, that donor-electron ionization-recombination terms are neglected in the right-hand side of Eqs.(3b,c), since we assume that local equilibrium is maintained and ionized donors mobility is sufficiently high. The terms in the right-hand-side of Eqs.(3b,c) should be obviously taken into account when the donors are immobile, but change their occupation degree dynamically in the electric potential created by the time-dependent electric voltage and mobile holes and electrons.

Boundary conditions to the kinetic constitutive equations are the following**.** The tip-surface interface is regarded impermeable for the donor ions, thus there is no ionic currents across the interface:

$$J_{dz}(\rho,0,t) = -D_d \frac{\partial}{\partial z}\delta N_d^+ - \eta_d \overline{N}_d^+ \frac{\partial}{\partial z}\varphi\bigg|_{z=0} = 0, \tag{4a}$$

Similarly, the fact that tip effect is localized yields $J_{dz}(\rho, z \to \infty, t) \to 0$. Since we regard the sample lateral surfaces placed in dielectric ambient, the circular currents are absent and thus:

$$\mathbf{J}_d(\rho \to \infty, t) = 0, \quad \mathbf{J}_n(\rho \to \infty, t) = 0, \tag{4b}$$

The boundary conditions for the electron current are taken in the linearized Chang-Jaffe form [53] as $J_{nz}(\rho, 0) = w(n(\rho, 0, t) - n_0)$. Hence

$$\left. D_n \frac{\partial}{\partial z}\delta n - \eta_n \bar{n} \frac{\partial}{\partial z}\varphi \right|_{z=0} = w\delta n(\rho, 0, t) \tag{4c}$$

The constant $w$ is positive rate constants related with the surface recombination velocity of electrons and holes, respectively.[54] The numerical values are determined by the electrode and MIEC material. If the rate constants are infinitely high, then the equilibrium electron concentrations at the contacts are pinned by the electrodes and independent of the applied voltage.[55] Thus the condition (4c) contains the continuous transition from the "open" ohmic contact ($w \to \infty \;\Rightarrow\; \delta n(\rho, 0, t) = 0$) to the interface limited kinetics ($w > 0$) and "completely blocking" contact ($w = 0$).

Using Fourier transformation in time ($t \to \omega$) and space $\{x, y\} \to \mathbf{k}$ domains, the steady state periodic solution of Eqs.(7)-(8) can be derived. For the semi-infinite sample it acquires the form:

$$\delta \tilde{N}_d^+(k, z, \omega) = \sum_{m=1,2} N_m(k, \omega)\exp(-s_m(k, \omega)z), \tag{5a}$$

$$\delta \tilde{n}(k, z, \omega) = \sum_{m=1,2}\left(1 - \frac{\varepsilon_0 \varepsilon}{\eta_d \bar{N}_d^+ q}\left(D_d\left(s_m^2(k, \omega) - k^2\right) - i\omega\right)\right)N_m(k, \omega)\exp(-s_m(k, \omega)z), \tag{5b}$$

$$\tilde{\varphi}(k, z, \omega) = \begin{pmatrix} \psi(k, \omega)\exp(-kz) + \\ \dfrac{1}{\eta_d \bar{N}_d^+}\displaystyle\sum_{m=1,2}\left(\frac{i\omega}{s_m^2(k, \omega) - k^2} - D_d\right)N_m(k, \omega)\exp(-s_m(k, \omega)z) \end{pmatrix}. \tag{5c}$$

Where $k = \sqrt{k_x^2 + k_y^2}$ and $\omega$ is the circular frequency of the voltage applied to the SPM tip. Cumbersome functions $N_m(k, \omega)$ and $\psi(k, \omega)$ are proportional to $\tilde{V}_0(k, \omega)$ and listed in **Appendix B.**

### III.2. Eigenvalues analysis

The dynamics of the ESM response can be understood from the frequency dependence of the positive eigenvalues $s_m(k,\omega)$ in Eqs. (5a-c). Under the electroneutrality condition, $\overline{N}_d^+ = \overline{n}$, for the system in the equilibrium and the Planck-Nernst-Einstein relation $\frac{\eta_d}{D_d} = \frac{\eta_n}{D_n} = \frac{q}{k_B T}$ (assumed hereinafter), the eigenvalues $s_{1,2}(k,\omega)$ have the form

$$s_{1,2}(k,\omega) = \sqrt{k^2 + \frac{i\omega}{2}\left(\frac{1}{D_n} + \frac{1}{D_d}\right) + \frac{1}{R_D^2} \pm \sqrt{\frac{1}{R_D^4} - \frac{\omega^2}{4}\left(\frac{1}{D_d} - \frac{1}{D_n}\right)^2}}, \qquad (6)$$

where $R_D = \sqrt{\frac{\varepsilon_0 \varepsilon D_d}{\eta_d \overline{N}_d^+ q}} = \sqrt{\frac{\varepsilon_0 \varepsilon D_n}{\eta_n \overline{n} q}} = \sqrt{\frac{\varepsilon_0 \varepsilon k_B T}{q^2 \overline{n}}}$ is the Debye screening radius and where $k_B$=1.3807×$10^{-23}$ J/K, $T$ is the absolute temperature. Note, that the screening radii corresponding to electrons and donors are the same because we consider the special case of MIEC with $\overline{N}_d^+ = \overline{n}$.

Equation (6) suggests that the eigenvalue spectrum contains two branches, $s_1(k,\omega)$ and $s_2(k,\omega)$ corresponding to the plus (minus) signs before the radical. Introducing the Maxwellian relaxation time, $\tau_M = \frac{(D_n + D_d)R_D^2}{2D_n D_d}$, the limiting cases of Eq. (6) are analyzed in the **Table 1**.

From the **Table 1** and Eq. (6), both eigenvalues have similar behavior at very low and very high frequencies. At high frequencies ($\omega\tau_M >> 1$) the asymptotic dispersion law is $s_{1,2}(k,\omega) = \sqrt{k^2 + i\omega / D_{d,n}}$, which we refer to "diffusive transfer" as governed by the diffusion coefficients $D_{d,n}$ only. At the same time, for low frequencies, $\omega\tau_M << 1$, the asymptotic dispersion law for $s_2(k,\omega) = \sqrt{k^2 + (2 + i\omega\tau_M)/R_D^2}$ is referred to as "drift transfer" (or, equivalently, electromigration). For $s_1(k,\omega)$ we still have drift-diffusion transfer as governed by the effective diffusion coefficient $D_{eff} = \frac{D_n + D_d}{2D_n D_d}$. Note, that in hypothetic degenerated case $D_n = D_d = D$ the diffusion charge transfer takes place for $s_1(k,\omega)$ in the entire frequency range, while $s_2(k,\omega)$ contains both drift and diffusion terms.

**Table 1.** Limiting charge transfer mechanism

| **Eigen values** | **Conditions on the material parameters and/or frequency range** | | |
|---|---|---|---|
| | Frequency $\omega\tau_M >> 1$ | Frequency $\omega\tau_M << 1$ | Arbitrary $\omega$, $D_n = D_d = D$ |
| $s_1(k,\omega)$ | $\sqrt{k^2 + \frac{i\omega}{D_d}}$ | $\sqrt{k^2 + i\omega\frac{\tau_M}{R_D^2}} \to k$ | $\sqrt{k^2 + \frac{i\omega}{D}}$ |
| $s_2(k,\omega)$ | $\sqrt{k^2 + \frac{i\omega}{D_n}}$ | $\sqrt{k^2 + \frac{2 + i\omega\tau_M}{R_D^2}}$ | $\sqrt{k^2 + \frac{i\omega}{D} + \frac{2}{R_D^2}}$ |
| **Classification of the charge transfer** | Diffusion transfer for electrons and ions | Mixed:<br>drift-diffusion transfer for $s_1$<br>drift transfer for $s_2$ | Mixed:<br>diffusion transfer for $s_1$<br>drift-diffusion transfer for $s_2$ |

Given that eigenvalues can be different, one should expect a transition region in which one of the eigenvalues has adopted the high-frequency asymptotic behavior, and the second one is still in the low frequency regime. Below we analyze the boundaries between the regimes and transition region as well as ESM responses in the regimes. We also note that Maxwell relaxation time $\tau_M \sim R_D^2$ is material constant that does not depend on the tip-surface geometry. Thus the screening radius $R_D$ can be the single length scale parameter only for the 1D case, which is realized for the planar top electrode and mathematically corresponds to the case $k = 0$. However, the system "semi-infinite MIEC + SPM tip" has two characteristic length scales – the tip characteristic size $R_0$ and the screening radius $R_D$. Consequently, the ionic diffusion time $\tau_D = R_0^2 / D_d$ is the second characteristic time for the system behavior. Below, we analyze these cases individually.

***I. Eigenvalues at k*=0.** To explore the possibility and to determine the boundaries between the regimes for $k = 0$, we calculated the dependence of the eigen values $s_{1,2}(0,\omega)$ vs. external voltage frequency $\omega$ from Eq.(6). For convenient comparison with previous studies, we normalize frequency on the characteristic diffusion time $\tau_D$ in **Figs. 2.** It is seen from **Figs. 2**, that the frequency dependences of $s_{1,2}(0,\omega)$ demonstrate two asymptotic frequency regimes for $\omega \leq \tau_M^{-1}$ and $\omega >> \tau_M^{-1}$, corresponding to purely diffusive and purely drift dynamics for both eigenvalues. In high frequency limit both characteristic eigenvalues tend to the diffusion limit $s_{1,2}(0,\omega) \to \sqrt{i\omega / D_{n,d}}$. Contrary, at the small frequencies, the behavior of $s_1$ and $s_2$ is different, while $s_1$ still demonstrate diffusion-like dependence, while $s_2$ behavior is determined by the screening radius $R_D$.

The degenerated case $D_d / D_n = 1$, where we have only the triple point, is shown in **Figs.2a,b**. For the case $D_d / D_n << 1$ the crossing of the roots real and imaginary parts, $\mathrm{Re}[s_{1,2}(0,\omega)]$ and $\mathrm{Im}[s_{1,2}(0,\omega)]$, occurs at frequency $\omega \approx \tau_M^{-1}$ (see dotted vertical line in plots **c-f**). Then $\mathrm{Im}[s_1(k,\omega)]$ decreases with the frequency increase in the range $\tau_M^{-1} < \omega < \omega_{ext}$. The extreme of $\mathrm{Im}[s_1(k,\omega)]$ is achieved at frequency $\omega_{ext} \approx \left| \frac{D_n + D_d}{D_n - D_d} \right| \frac{\sqrt{D_n D_d}}{2R_D^2}$ (see circle at the red curves in plots **c-f**). In the intermediate frequency range, $\tau_M^{-1} < \omega < \omega_{\max}$ (where $\omega_{\max} \approx \frac{10}{R_D^2} \max[D_n, D_d]$), the root $s_2(k,\omega)$ is diffusive, while other $s_1(k,\omega)$ exhibits mixed drift-diffusion behavior. Then imaginary parts of both eigenvalues increase with the further frequency increase. The frequency dependence becomes linear in log-log scale at $\omega > \omega_{\max}$. Finally, condition $D_n << D_d$ (i.e. $\eta_n << \eta_d$) seems unrealistic, but can be analyzed similarly to the case discussed above.

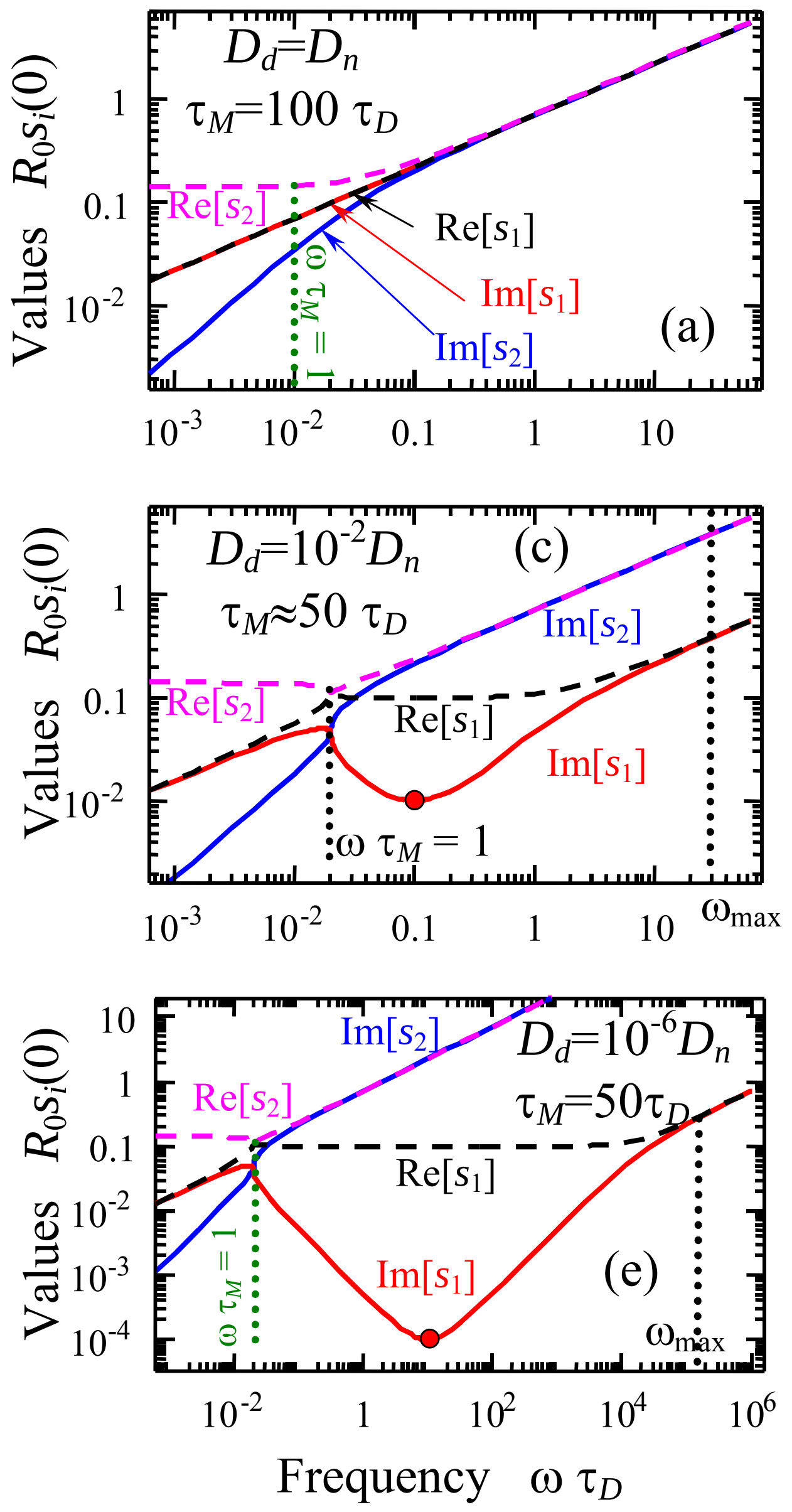

**Fig. 2.** The real (Re), imaginary (Im) parts of the eigenvalues $s_{1,2}(0,\omega)$ vs. external field frequency ω calculated from Eq.(6) for $D_d/D_n = 1$ (a, b) and $10^{-2}$ (c, d) and $10^{-6}$ (e, f); $R_D/R_0 = 10$ (a, c, e) and 0.1 (b, d, f).

***II. Eigenvalues at k=1/R*₀*.*** The value $k=1/R_0$ reflects the tip size influence on the response depth. Direct calculations show, that the parameter $R_D^2/R_0^2$ is approximately equal the ratio of Maxwell relaxation time to the diffusion time $2\tau_M/\tau_D$, once $D_n >> D_d$. This follows directly from estimate $\frac{\tau_M}{\tau_D} \equiv \frac{D_d}{R_0^2} \cdot \frac{(D_n + D_d) R_D^2}{2 D_n D_d} = \frac{R_D^2}{R_0^2} \cdot \frac{(D_n + D_d)}{2 D_n} \underset{D_n >> D_d}{\approx} \frac{R_D^2}{2 R_0^2}$. Thus, the relation between the diffusion time $\tau_D$ and Maxwell relaxation time $\tau_M$ should determine the features of response behavior, as illustrated for the frequency dependence of $s_{1,2}(R_0^{-1}, \omega)$ in **Figs. 3**. For convenient comparison with previous studies, we normalize frequency on the characteristic diffusion time $\tau_D$**.** The degenerated case $D_d / D_n = 1$ are shown in **Figs.3a,b,** the realistic cases $D_d / D_n << 1$ are shown in **Figs. 3c-f.**

**Figure 3** illustrates that for nonzero values of $k = R_0^{-1}$ the spectrum of eigenvalues $s_{1,2}(k, \omega)$ reveal characteristic behaviour at low frequencies $\omega < \tau_M^{-1}$: the real parts $\mathrm{Re}[s_1(k, \omega)]$ and $\mathrm{Re}[s_2(k, \omega)]$ tend to the constant value (not to zero as for $k = 0$), while the imaginary parts $\mathrm{Im}[s_1(k, \omega)]$ and $\mathrm{Im}[s_2(k, \omega)]$ tend to zero in the limit $\omega \to 0$. Hence, both eigenvalues $s_{1,2}(k > 0, \omega)$ have nonzero values in the limit $\omega \to 0$, while both eigenvalues real and imaginary parts split.

In other words, $\mathrm{Im}[s_{1,2}(k, \omega)]$ is only weakly affected by $k$ value, while this is not the case for the real parts. Actually, the crossing of the eigenvalues imaginary parts, $\mathrm{Im}[s_2(0, \omega)]$ and $\mathrm{Im}[s_1(0, \omega)]$, happens at frequency $\omega \approx \tau_M^{-1}$ (see dotted vertical line in **Figs. 2c-f** and compare with **Figs. 2c-f**). Then $\mathrm{Im}[s_1(k, \omega)]$ decreases with the frequency increase in the range $\tau_M^{-1} < \omega < \omega_{ext}$. The extremum of $\mathrm{Im}[s_1(k, \omega)]$ is achieved at frequency $\omega_{ext} \approx \left| \frac{D_n + D_d}{D_n - D_d} \right| \frac{\sqrt{D_n D_d}}{2 R_D^2}$ (see circle at the red curves in **Figs. 3c-f** and compare its position with the one from **Figs. 2c-f**). Then imaginary parts of both eigenvalues increase with the frequency increase for frequencies $\omega > \omega_{ext}$. The frequency dependence becomes linear in log-log scale at $\omega > \omega_{\max}$ for $s_2(k, \omega)$ and $\omega_{\max} \approx \frac{10}{R_D^2} \max[D_n, D_d]$, similarly to the case of $k = 0$.

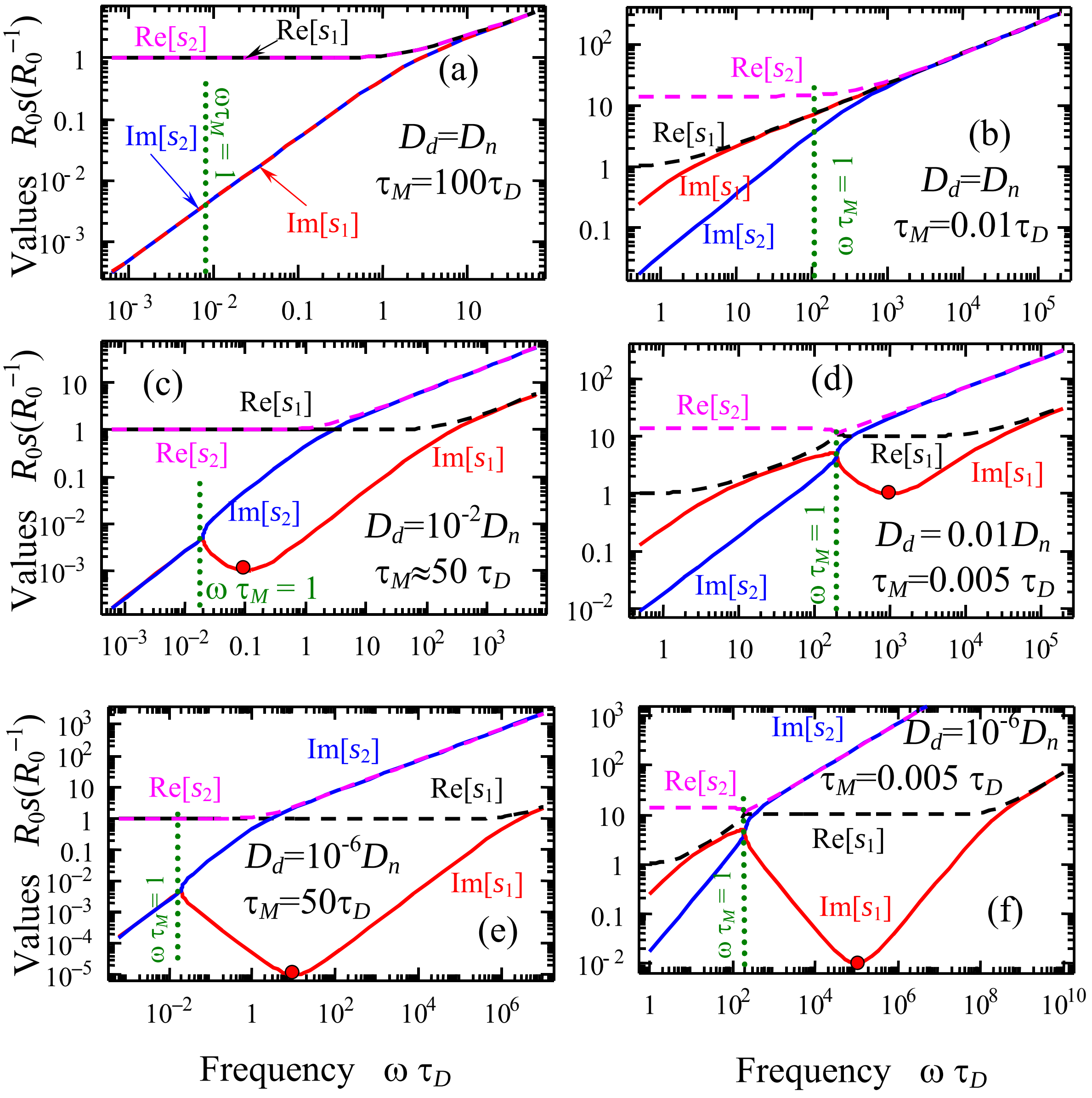

**Fig. 3.** The real (Re), imaginary (Im) parts of the eigenvalues $s_{1,2}\left(R_0^{-1},\omega\right)$ vs. external field frequency ω calculated from Eq.(6) for $D_d/D_n=1$ (a, b) and $10^{-2}$ (c, d) and $10^{-6}$ (e, f); $R_D/R_0=10$ (a, c, e) and 0.1 (b, d, f).

### III.3. Mechanical surface displacements in ESM

The problem of mechanical stresses developing in the electrochemical systems have been recently addressed by a number of authors, including both the cases of macroscopic material and case of spherical particle[56, 57, 58, 43]. For the latter, both decoupled and coupled [57, [58, 40, 41] numerical solutions are available. Importantly, the error induced by decoupling approximation is shown to be proportional to the square of the molar expansion tensor and generally does not exceed 30%, well below the uncertainty of tip-surface contact radius in a typical SPM experiment.

For the particular case when the chemical contribution is the dominant active mechanisms for strain, the equations of state (Hooke's law for the chemically active solid) for isotropic elastic media, subjected to the ionic flux relates concentration excess $\delta N_d^+(\mathbf{r},t)$, mechanical stress tensor $\sigma_{ij}$ and elastic strain $u_{ij}$ are the following [43, 57]:

$$u_{ij}(\mathbf{r},t)=\beta_{ij}\ \delta N_d^+(\mathbf{r},t)+s_{ijkl}\sigma_{kl}(\mathbf{r},t). \tag{7}$$

Here $s_{ijkl}$ is the tensor of elastic compliances, $\beta_{ij}$ is the Vegard expansion tensor. The Vegard contribution (chemical expansion) $\beta_{ij}\ \delta N_d^+(\mathbf{r},t)$ describes the lattice deformations under the small changes of composition $\delta N_d^+(\mathbf{r},t)=\left(N_d^+(\mathbf{r},t)-\overline{N}_d^+\right)$. We further restrict the analysis to the transversally isotropic Vegard tensor $\beta_{ij}=\delta_{ij}\beta_{ii}$ (with $\beta_{11}=\beta_{22}\neq\beta_{33}$)

In subsequent analysis we note that the typical contact area in SPM experiment is well below micron-scale. The corresponding intrinsic resonance frequencies of material are thus in the GHz range, well above the practically important limits both in terms of ion dynamic, and SPM-based detection of localized mechanical vibrations. As described in Ref.[43] this allows using quazistatic approximation for mechanical phenomena. Here, we solved the general equation of mechanical equilibrium in the quasi-static case that leads to the equation for mechanical displacement vector $u_i$ in the bulk of the system with appropriate boundary conditions $\sigma_{3j}(z=0,t)=0$ on the free surface $z=0$. The surface displacement at the tip-surface junction $z=0$ induced by the redistribution of mobile donors, as detected by SPM electronics, for elastically isotropic semi-space is:

$$\begin{aligned} u_3(\rho,0,\omega) &= \int_0^\infty dkJ_0(k\rho)k\int_0^\infty dz\left(\beta_{33}(1+kz)+\beta_{11}(1+2\nu-kz)\right)\exp(-kz)\delta N_d^+(k,z,\omega) \\ &= \int_0^\infty dkJ_0(k\rho)k\sum_{m=1,2}N_m(k,\omega)\left(\beta_{33}\frac{2k+s_m}{(k+s_m)^2}+\beta_{11}\frac{2k\nu+s_m(1+2\nu)}{(k+s_m)^2}\right) \end{aligned} \tag{8}$$

Here $\nu$ is the Poisson coefficient, $k^2 = k_x^2 + k_y^2$, $\delta N_d^+(k,z,\omega)$ is the 2D-Fourier image and frequency spectrum of the ion concentration field variation $\delta N_d^+(\mathbf{r},t)$, and $J_0(x)$ is the Bessel function of zero order.

From Eq.(8), the ESM response depends not only on the eigenvalues $s_m(k,\omega)$, but also on the coefficients $N_m(k,\omega)$ and denominators $(k+s_m)^{-2}$. After the integration on the wave vector $dk$, a characteristic value of $k$ determined by the scale of external electric field (e.g. $k \sim 1/R_0$ ) appeared in the final expression. While unambiguous conclusions on the ESM response spectra cannot be derived based on the analyses of the $s_m(k,\omega)$ spectra alone, it allows for semiquantitative analysis of the frequency dependent electromechanical response.

Here, we define the electromechanical response of mixed electronic-ionic conductors as purely ionic for the situation when the electronic mobility is much higher then ionic, $D_n >> D_d$ (allowing for the Nerst-Einstein relation). In this case, the electron charge variations in Eqs.(3a,b) can be neglected, i.e. one could regard that $\left|\delta N_d^+\right| >> \left|\delta n\right|$. Note, that this assumption does not violate the conservation laws, but implies that the electrons move much faster and play the role of instantly responding screening charge.

Calculations of the **purely ionic response** are summarized in **Appendix B.2**. Corresponding solution for the ESM response is:

$$u_3(\rho,\omega) = \frac{q\overline{N}_d^+}{k_B T}\int_0^{\infty} dk k^2 \widetilde{V}_0(k,\omega) J_0(k\rho) \frac{\beta_{33}(2k + s(k,\omega)) + \beta_{11}(2k\nu + s(k,\omega)(1+2\nu))}{(k + s(k,\omega))\left(k^2 + k\cdot s(k,\omega) + \dfrac{i\omega}{D_d}\right)}, \quad (9a)$$

For the case of isotropic Vegard tensor, $\beta_{11} = \beta_{33}$, Eq.(9a) is reduced to

$$u_3(\rho,\omega) = \frac{q\overline{N}_d^+}{k_B T}\int_0^{\infty} dk \frac{2(1+\nu)\beta k^2 \widetilde{V}_0(k,\omega) J_0(k\rho)}{k^2 + k\cdot s(k,\omega) + \dfrac{i\omega}{D_d}} \quad (9b)$$

Where the characteristic eigenvalue $s(k,\omega)$ is introduced as

$$s(k,\omega) = \sqrt{k^2 + \frac{i\omega}{D_d} + \frac{1}{R_D^2}}. \quad (9c)$$

Note, that the last term $R_D^{-2}$ in the expression for $s(k,\omega)$ describes the role of carriers electromigration that was neglected in our previous study of diffusionally-coupled ESM response (compare Eq.(9c) with Eq.(13) from Ref.[ 43]).

The frequency-dependent strain signal in ESM provides an analog of classical current-based electrochemical impedance spectroscopy. For numerical estimations the Fourier image of the surface potential $V_0(\rho,t)$ can be taken as $\widetilde{V}_0(k,\omega)=V(\omega)R_0\frac{J_1(kR_0)}{k}$, where $J_1(x)$ is the Bessel function of the first order, $R_0$ is the tip-surface contact radius.

The important limiting cases of Eq. (9) are the (a) ***local ESM response*** defined as the surface displacement in the point **r** = 0, (b) ***averaged ESM response*** defined as the surface displacement averaged by the contact area $\langle u_3(\rho,\omega)\rangle=\frac{2}{R_0^2}\int_0^{R_0}u_3(\rho,\omega)\rho d\rho$, (c) ***ESM response*** at the contact line $\overline{u}_L(\omega)=u_3(R_0,\omega)$. The local ESM response provides the first-order approximation for ESM signal in general, and is well suited for the cases when the electrostatic tip radius is much larger then mechanical contact radius (e.g. water droplet effects). Cases (b) and (c) provide the input for the approximate contact mechanical solutions for the different tip geometries. Analytical expressions for these responses for the case $\beta_{ii}=\beta$ are listed in the **Table 2**, which specifically indicates the limiting cases for low frequency and high frequency.

Let us underline, that obtained analytical results show that dielectric constant$\varepsilon$ influence on the Debye screening radius as $R_D=\sqrt{\frac{\varepsilon_0\varepsilon k_B T}{q^2\overline{n}}}\sim\sqrt{\varepsilon}$ and Maxwellian relaxation time $\tau_M=\frac{(D_n+D_d)R_D^2}{2D_nD_d}\sim\varepsilon$. Both parameters $R_D$ and $\tau_M$ increase with $\varepsilon$ increase. The screening radius and relaxation time determine the relative contribution of the electromigration into the ESM response characteristics (see Eqs.(9)). Since the relative contribution of the electromigration into the ESM response decreases with $R_D$ increase and $\tau_M$ increase, $\varepsilon$ increase leads to decrease of the electromigration impact. Diffusion contribution is virtually independent on the dielectric permittivity.

**Table 2.** Limiting frequency regimes for the ionic ESM response

| Frequency range | Maximal ESM response, $u_3(0,\omega)$ | Average ESM response $\langle u_3(\rho,\omega)\rangle = \frac{2}{R_0^2}\int_0^{R_0} u_3(\rho,\omega)\rho d\rho$ | ESM response at the contact line $\bar{u}_L(\omega) = u_3(R_0,\omega)$ |
|---|---|---|---|
| $\omega\tau_D << 1$ | $2(1+\nu)\beta V(\omega)\times \times\frac{R_D R_0}{(2R_D+R_0)}\frac{q\bar{N}_d^+}{k_B T}$ | $\frac{2(1+\nu)\beta V(\omega) R_D R_0}{\frac{3\pi}{4}R_D + R_0}\frac{q\bar{N}_d^+}{k_B T}$ | $2(1+\nu)\beta\bar{N}_d^+\frac{qV(\omega)}{k_B T}\times\left(\frac{R_0 R_D}{2R_0+\pi R_D}\right)$ |
| $\omega\tau_D >> 1$ | $\frac{2(1+\nu)\beta V(\omega)}{\sqrt{\frac{i\omega}{D_d}}\left(1+R_0\sqrt{\frac{i\omega}{D_d}}\right)}\frac{q\bar{N}_d^+}{k_B T} \sim \frac{1}{i\omega\tau_D}$ | $2(1+\nu)\beta V(\omega)\frac{q\bar{N}_d^+}{k_B T}\frac{D_d}{i\omega R_0\pi}\times \times\left(2(\gamma-2)+\ln\left(2^5\frac{i\omega R_0^2}{D_d}\right)\right)$ * | $2(1+\nu)V(\omega)\beta\frac{q\bar{N}_d^+}{k_B T}\frac{D_d}{i\omega R_0 4\pi}\times\left(2(\gamma-1)+\ln\left(2^5\frac{i\omega R_0^2}{D_d}\right)\right)$ * |

*Here, $\gamma = 0.577216\ldots$ is Euler's constant, and $\tau_D = R_0^2/D_d$ as introduced above.

The frequency spectra of ESM response $u_3(\mathbf{r}=0,\omega)$ amplitude and phase are shown in **Figs. 4** and **5** correspondingly for isotropic of Vegard tensor, $\beta_{ii}=\beta$, and ***three dimensionless parameters***: diffusion coefficients ratio $D_d/D_n$, electronic subsystem parameter $wR_0/D_n$ and the ratio of the screening radius to the tip effective size, $R_D/R_0$**.** Similarly to **Figs.2,3** we normalize frequency on the characteristic diffusion time $\tau_D$ in **Figs. 4-5,** since it is the natural scale of the response spectra**.** Dotted curves in **Figs.4-5** represent the results for purely ionic response given by Eqs.(9), while dashed line is the high frequency diffusion transfer limit $u_3 \sim 1/(i\omega\tau_D)$ (see the last raw of the **Table 2**).

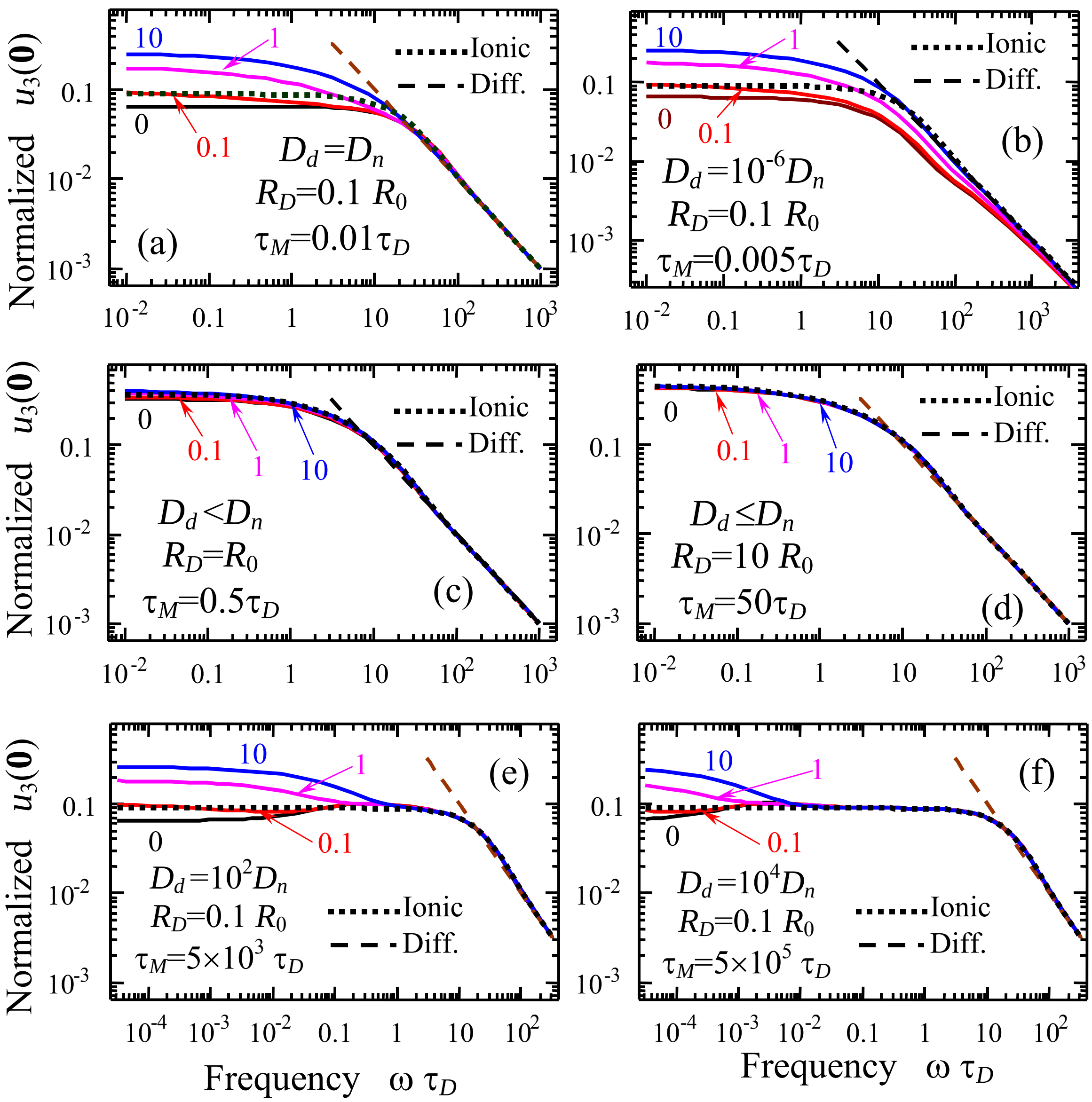

**Fig. 4.** The amplitude of the ionic response vs. external field frequency ω calculated from Eq.(8) for $D_d/D_n=1$ (a), $\le 10^{-6}$ (b), ≤1 (c, d), $10^2$ (e) and $10^4$ (f); $R_D/R_0=0.1$ (a, b, e, f) 1 (c), 10 (d) and different values of parameter $wR_0/D_n$ =0, 0.1, 1, 10 (numbers near the solid curves). Dotted curve represent the results of the model with purely ionic response, while dashed line is for high frequency diffusion limit $u_3\sim 1/(i\omega\tau_D)$. The response is normalized by $R_0(1+\nu)\beta\overline{N}_d^+\eta_d V(\omega)/D_d$

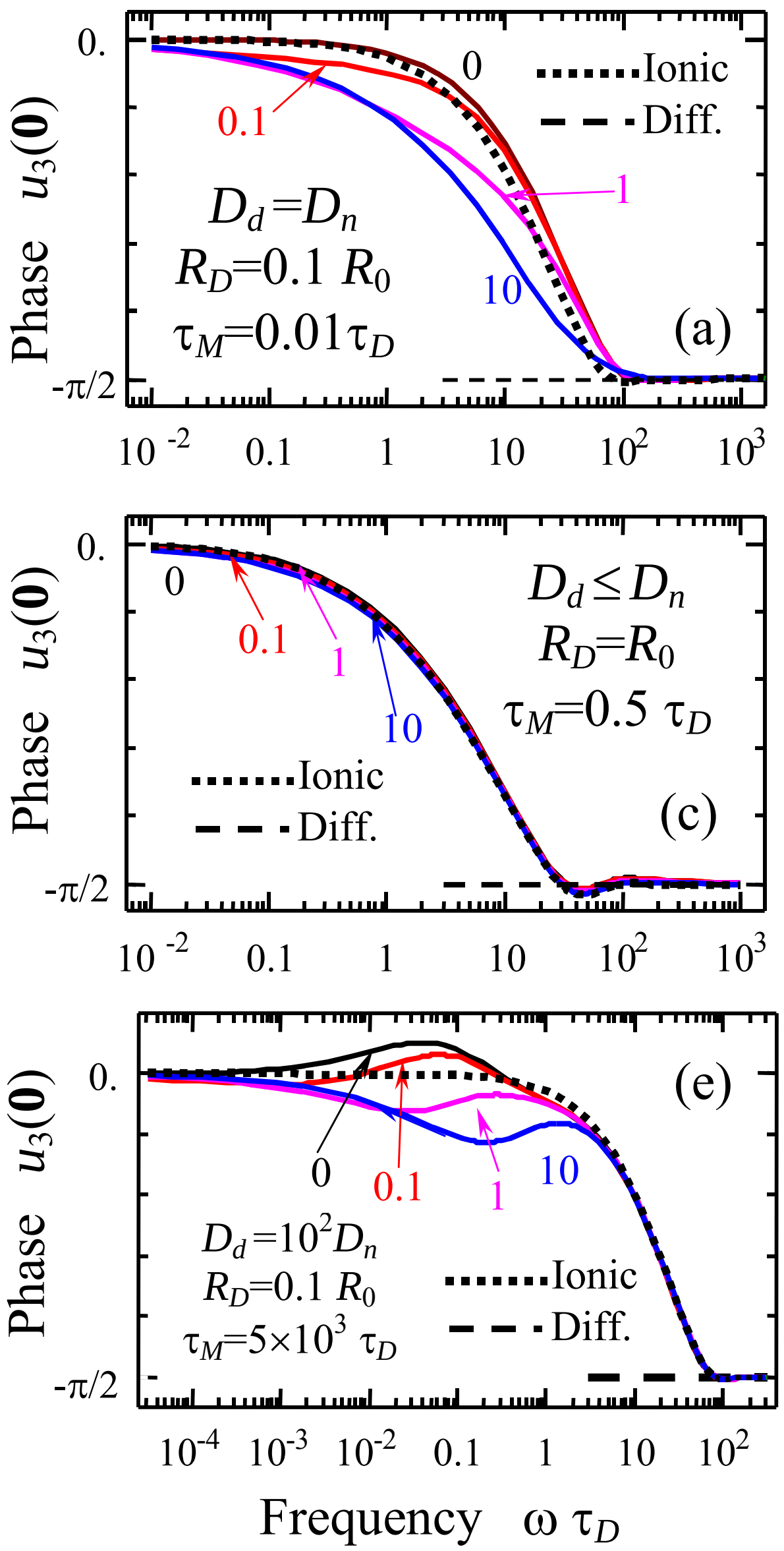

**Fig. 5.** The phase of the ionic response vs. external field frequency ω calculated from Eq.(8) for $D_d/D_n = 1$ (a), $\leq 10^{-6}$ (b), $\leq 1$ (c, d), $10^2$ (e) and $10^4$ (f); $R_D/R_0 = 0.1$ (a, b, e, f) 1 (c), 10 (d) and different values of parameter $wR_0/D_n$ =0, 0.1, 1, 10 (numbers near the solid curves). Dotted curve represent the results of the model with purely ionic response, while dashed line is for high frequency diffusion limit $u_3 \sim 1/(i\omega\tau_D)$. The response is normalized by $R_0(1+\nu)\beta\overline{N}_d^+\eta_d V(\omega)/D_d$ .

This analysis allows the influence of the parameters $D_d/D_n$, $wR_0/D_n$ and $R_D/R_0$ on the frequency-dependent electromechanical response to be determined. In particlar the parameter $R_D/R_0$ is the primary factor affecting the response, since electric field distribution in material below the tip mainly determines the ESM response spectra shape. Correspondingly, the ratio $D_d/D_n$ and parameter $wR_0/D_n$ have only weak influence on the frequency-dependent ESM response.

Note that the strong inequality $R_D >> R_0$ corresponds to the very week screening of the tip inhomogeneous electric field in the response region $r \le R_0$, while the inequality $R_D \le R_0$ and especially $R_D << R_0$ correspond to the strong screening of the electric field in the response region. Only under the conditions $R_D << R_0$ will the ESM response amplitude and phase depend on the on the parameter $wR_0/D_n$ (i.e. on recombination rate $w$ and electron diffusion coefficient $D_n$) in the relatively low frequencies range $\omega\tau_D \lesssim (1-10)$.

In the high frequency limit $\omega\tau_D >> 10$ the ESM response $u_3 \sim 1/(i\omega\tau_D)$, and its amplitude and phase are weakly dependent on $R_D/R_0$, $D_d/D_n$ and $wR_0/D_n$ for any set of parameters, since $\tau_D = R_0^2/D_d$ is the primary parameter determining system behavior (see curves tending to the dashed line in **Fig. 4** and **5** at high frequencies). This behavior can be readily explained by the fact that the diffusion length $\sqrt{\omega/D}$ becomes much smaller than the screening radius $R_D$ for high frequencies. In the case $R_D^2 \ge R_0^2$ (i.e. tip is small compared to the Debye length) the electronic subsystem parameter $wR_0/D_n$ practically does not affect the ESM response (see **Fig. 4 c,d** and **5c,d**).

Moreover, we note that the spectrum of $u_3$ tends to that of "pure" ionic system with $R_D/R_0$ increase (dotted curves coincide with solid ones in **Fig. 4 c,d** and **5c,d**). Thus, the results given by Eqs.(9) (dotted curves) and the limiting cases listed in the **Table 2** works well under the condition $R_D/R_0 > 1$ with accuracy increasing with the ratio $R_D/R_0$. An analysis of the limits for the purely ionic response $u_3(0,\omega)$ (see **Table 2**) gives the approximate expression for the cut-off frequency $\omega_0 \approx \frac{1}{\tau_D}\left(2+\frac{R_0}{R_D}\right)$ that scales with $\tau_D$ and depends only on the ratio $R_D/R_0$ (see next subsection).

It is seen from the **Fig. 4 a-d** and **5a-d** that approximation (9) is rather accurate for the case $D_d/D_n \leq 1$ (compare the curves calculated at $D_n/D_d = 10^{-6}$ and the ones calculated at $D_n/D_d = 1$). Only in the hypothetical case $D_n << D_d$ (shown for comparison only in **Fig. 4 e, f** and **5e, f**) the response is dominated by the carrier diffusion, except the low frequencies $\omega\tau_D \lesssim D_n/D_d$, where the contribution of electronic subsystem is apparent.

### III.4. ESM response spectra shape: static response and cut-off frequency

Direct comparison of electromigration and diffusion contributions into ESM response of MIECs can be performed with the help of the schematics shown **Figs. 6**. Here we approximate the ESM response spectra in the log-log scale by a rectangular trapezoid with height $u_0$, the smallest basis $\omega_0$ (cut-off frequency) and the angle $\pi/4$, since the high frequency diffusion limit is $u_3 \sim D_d/\left(i\omega R_0^2\right)$. Using the trapezoid approximation the ESM response has 3 regions:

(a) drift (or electromigration) region at $\omega << \omega_0$, where the response is almost constant $u_3 \approx u_0$;

(b) diffusion region at $\omega >> \omega_0$, where the response vanishes with frequency increase as $u_3 \sim D_d/\left(i\omega R_0^2\right)$;

(c) drift&diffusion crossover region with smeared boundaries located at the vicinity of the cut-off frequency $\omega_0$, e.g. the $0.3\omega_0 < \omega < 3\omega_0$.

An analysis of the limits for the purely ionic response (see **Table 2**) gives the approximate expressions

$$u_0 \approx 2(1+\nu)\frac{\beta \overline{N}_d^+ \cdot R_D R_0}{\left(2R_D + R_0\right)}\frac{qV}{k_B T} \text{ and } \omega_0 \approx \frac{1}{\tau_D}\left(2+\frac{R_0}{R_D}\right) = \frac{D_d\left(2R_D + R_0\right)}{R_D R_0^2}. \qquad (10)$$

However, the realistic mixed response appeared much more complex.

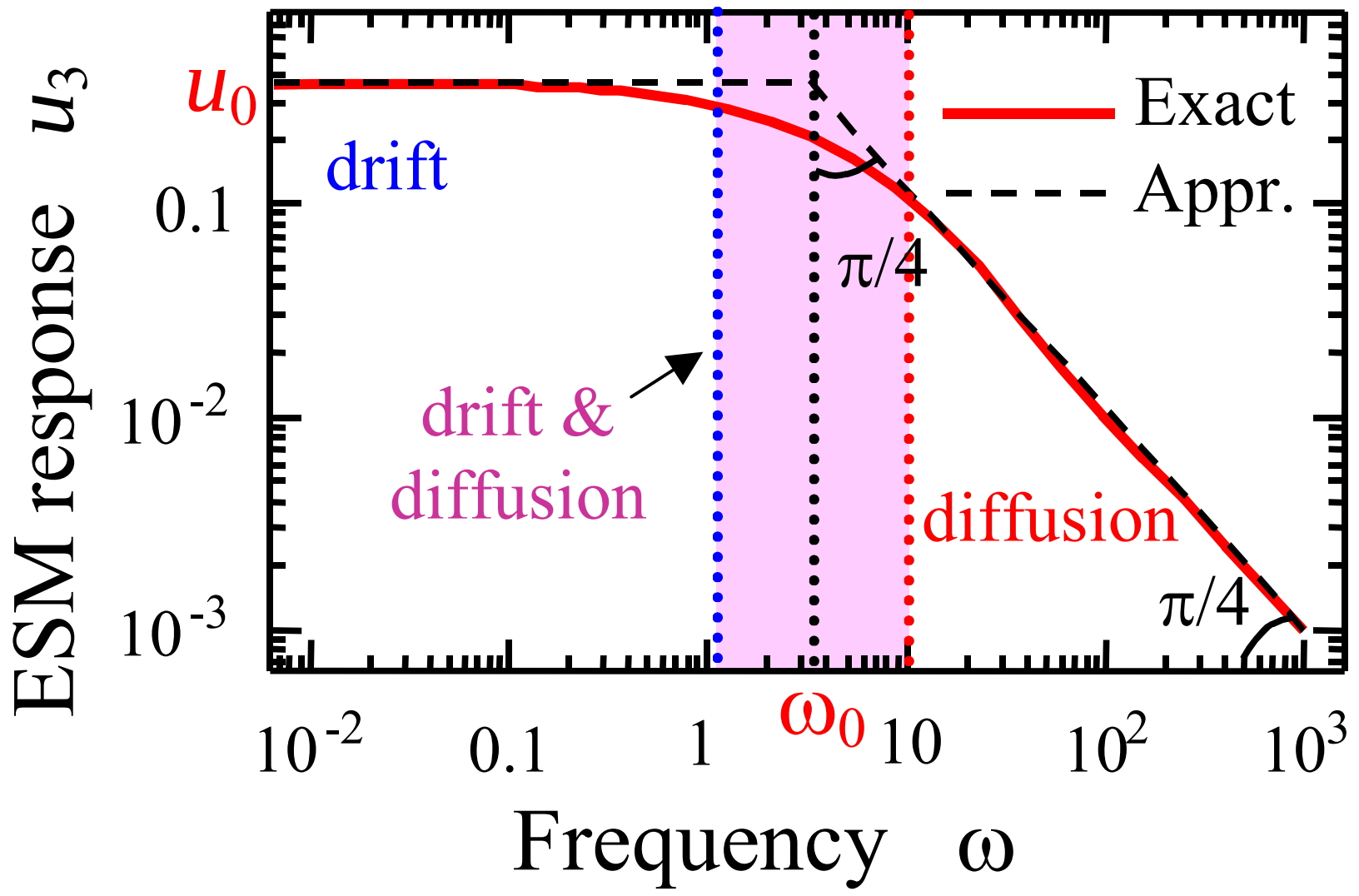

**Fig. 6.** Schematics of the ESM response spectra in the log-log scale. Solid curve is an exact expression, dashed curve is the approximation by rectangular trapezoid with height $u_0$, the smallest basis $\omega_0$ and the angle $\pi/4$. The drift, drift&diffusion and diffusion regions are shown.

### III.5. Dependence of ESM response on the tip effective size (or contact radius)

Since the approximation (9) was shown to work for most of the realistic cases at least semi-quantitatively, the dependence of the purely ionic ESM responses $u_3(0,\omega)$, $\langle \overline{u}_3(\rho,\omega)\rangle$ and $u_3(R_0,\omega)$ on the tip effective size $R_0/R_D$ was calculated from Eqs.(9). Results are shown in **Fig. 7** for several values of the frequency ω.

ESM responses shown in **Fig. 7** have maxima at approximately the same ratio $R_{\max}/R_D$, where the value $R_{\max}$ depends on the ratio $R_D^2\omega/D_d$. The maximum appears, since ESM response increases with $R_0$ for frequencies $\omega < D_d/R_0^2$, while it drops as $\sim R_0^{-1}$ at $\omega >> D_d/R_0^2$ (see **Table 2**). Note, that the position of the maxima $R_{\max}$ shifts to the higher values of $R_0$ with the frequency decrease. At very low frequencies and in the static limit (ω=0, dotted curves) the response grows linearly with the tip size and saturates when $R_0$ overcomes the screening radius $R_D$. More rigorously, for frequencies $R_D^2\omega/D_d \le 10^{-3}$ and tip radii within the range $R_D << R_0 < D_d/(\omega R_D)$ the response becomes almost independent on the tip (or contact) radius $R_0$. This analysis suggest that generally ESM response with ion-blocking electrodes is dependent on the contact radius, while ESM in the diffusion-coupled regime as analyzed in Ref.[43] as well as response in Piezoresponse force microscopy (PFM) are almost independent on the contact radius [59, 60].

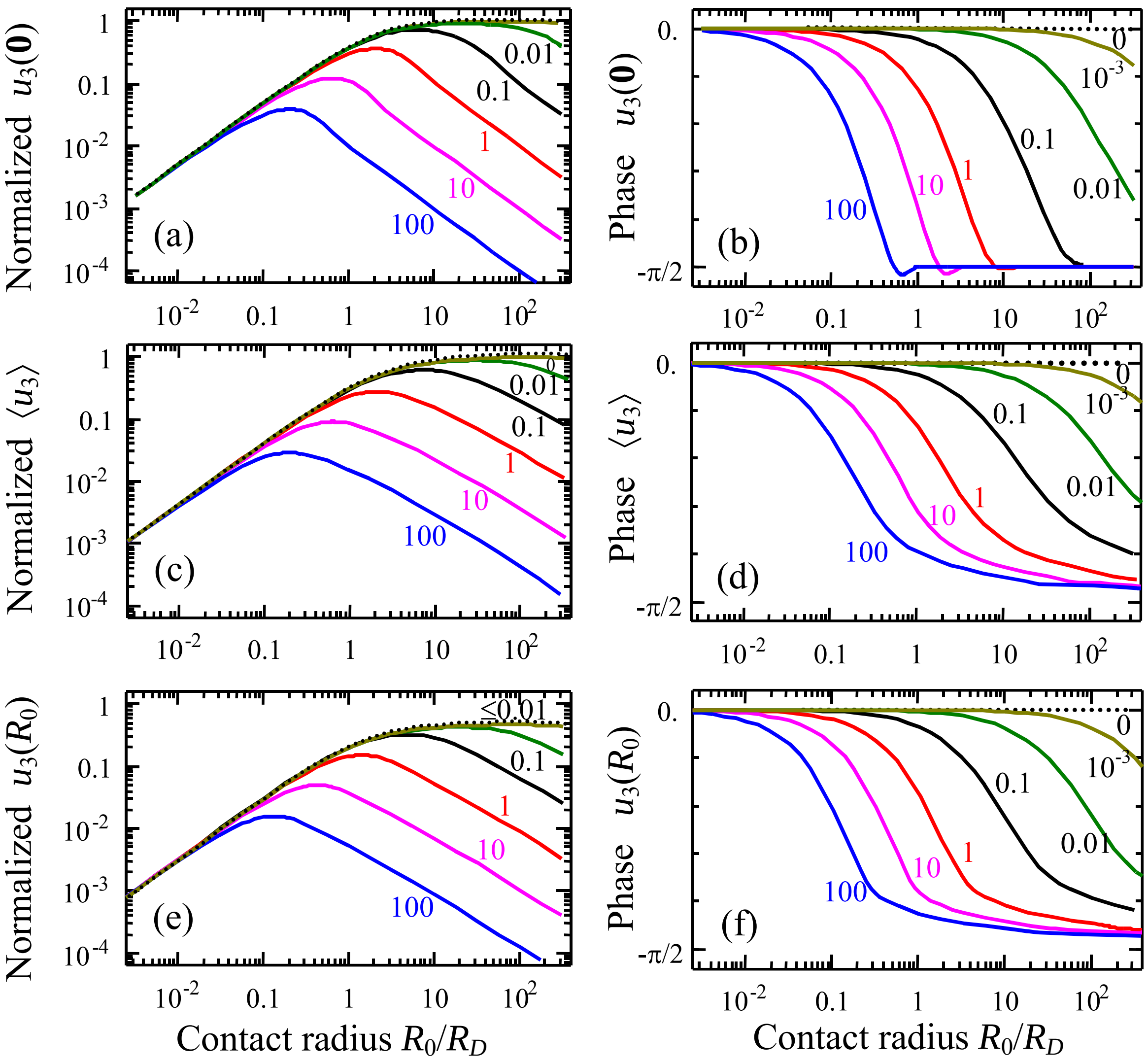

**Fig. 7.** The amplitude (a, c, d) and phase (b, d, f) of the **purely ionic response** vs. tip effective size $R_0/R_D$ for several values of external field frequency $R_D^2\omega/D_d$ =0, $10^{-3}$, 0.01, 0.1, 1, 10, 100 (numbers near the curves). The response maximal value at ρ=0 (a, b), averaged on the tip area (c, d) and averaged on the tip perimeter. The response is normalized on the combination $R_D(1+\nu)\beta\bar{N}_d^+\eta_d V(\omega)/D_d$ .

## IV. Discussion

In this section, we analyze the ESM mechanism for several ionic materials, and compare it with the results for purely reaction-driven case .

### IV.1. ESM response for Li-containing ionic materials

The amplitude and phase of the ESM response vs. external field linear frequency $f$ ($\omega = 2\pi f$) calculated for the material parameters of $LiCoO_2$ and $LiMn_2O_4$ are shown in **Fig. 8** for different ratio $R_0/R_D = 0.5$, 2, 10, 50. For the case when the contact radius $R_0$ is much larger then the Debye length $R_D$ and $D_d/D_n << 1$ the mixed response is much higher than the purely ionic one at low frequencies (compare dotted curves "50", "10" calculated as purely ionic response from Eqs.(9) with solid curves "50", "10" calculated as mixed response from Eqs.(8)). For the case $R_0/R_D \leq 2$ the mixed response is very close to the ionic one (compare dotted curves "2", "0.5" with solid curves "2", "0.5". $LiCoO_2$ and $LiMn_2O_4$ ESM response depends on the electronic subsystem parameters ($w$ and $D_n$) only for the case $R_0/R_D \geq 10$. These results are in agreement with analyses of section **III.3**.

Note that the ESM response amplitude strongly decreases with the Debye length increase: in order to obtain detectable values ~ (1-10) pm at low frequencies, the Debye length should be not smaller that 1-2 nm. On the other hand the values $R_D$ ~ 1 nm correspond to the realistic concentration of ionized donor atoms.

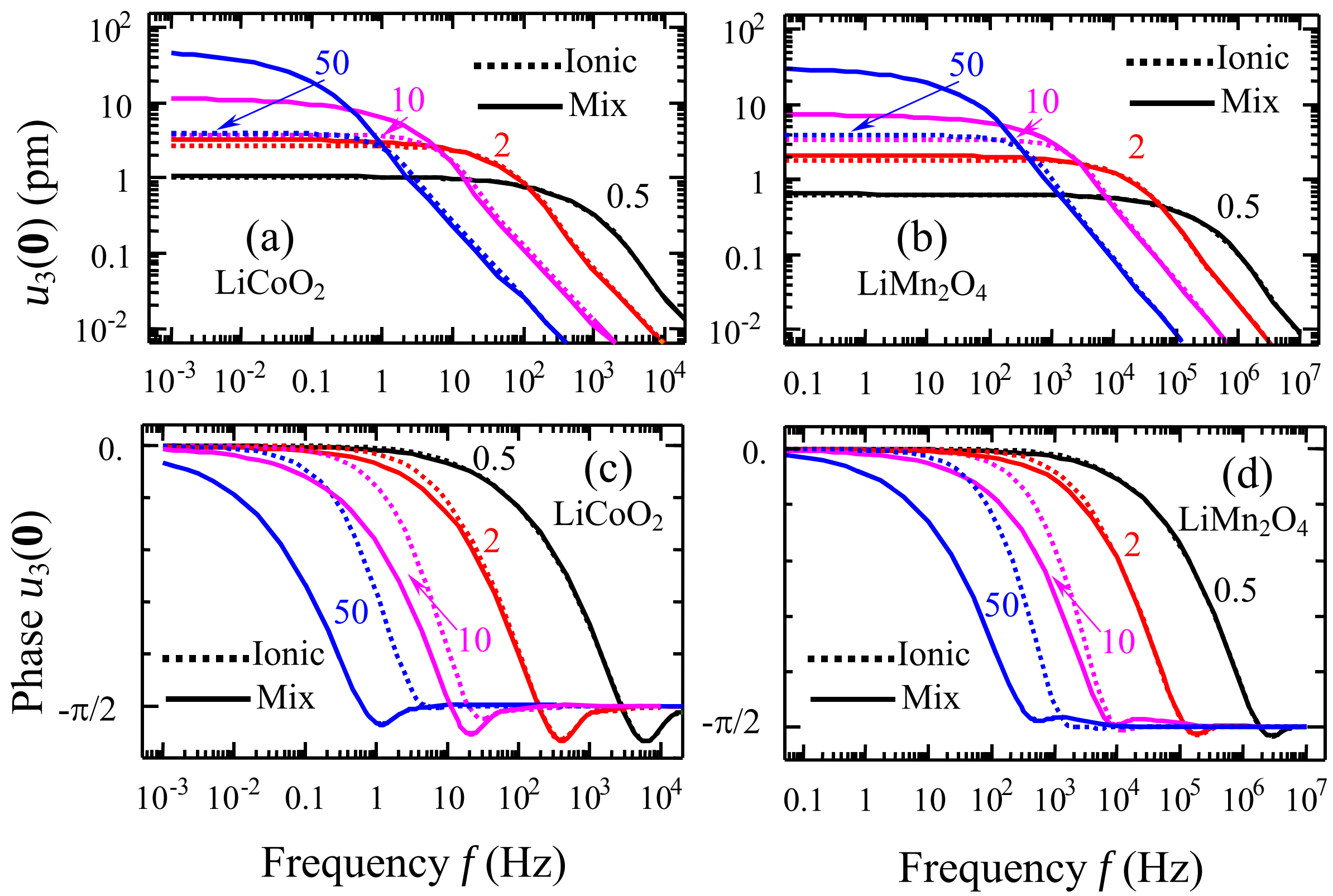

**Fig. 8.** The amplitude (a, b) and phase (c, d) of the ESM response vs. external field frequency $f$ calculated for the material parameters of $LiCoO_2$ (a, c) and $LiMn_2O_4$ (b, d) for different $R_0$=0.5, 2, 10, 50 nm (numbers near the curves) and fixed Debye length $R_D \approx 1$ nm. Dotted curves are purely ionic response calculated from Eqs.(9), solid curves are mixed response calculated from Eqs.(8) for diffusion coefficients ratio $D_d / D_n = 10^{-2}$ and rate constant $wR_0 / D_n = 5$. Material parameters of $LiCoO_2$ and $LiMn_2O_4$ are listed in the **Table 3**.

**Table 3**. Material parameters used in calculations

| **Composition and Refs.** | $D_d$ **(cm²/s)** | ν | $\beta_{11}=\beta_{22}$ **($10^{-30}$ m³)** | $\beta_{33}$ **($10^{-30}$ m³)** | $n=N_d^+$ **($10^{25}$ m⁻³)** | ε | $R_D$ **(nm)** |
|---|---|---|---|---|---|---|---|
| $LiCoO_2$ [61, 62, 63] | $2.4\times10^{-12}$ | 0.27 | <0.3 | 2.50 | 3.11 | ~ 10 | 1.2 |
| $LiMn_2O_4$ [64] | $1.0\times10^{-9}$ | 0.33 | 1.96 | 1.96 | 1.38 | ~ 10 | 1.4 |

The dependences of the static response $u_0$ and cut-off frequency $\omega_0$ (introduced in the section **III.4**) on the $LiCoO_2$ and $LiMn_2O_4$ material parameters were calculated numerically and shown in **Figs.9.**

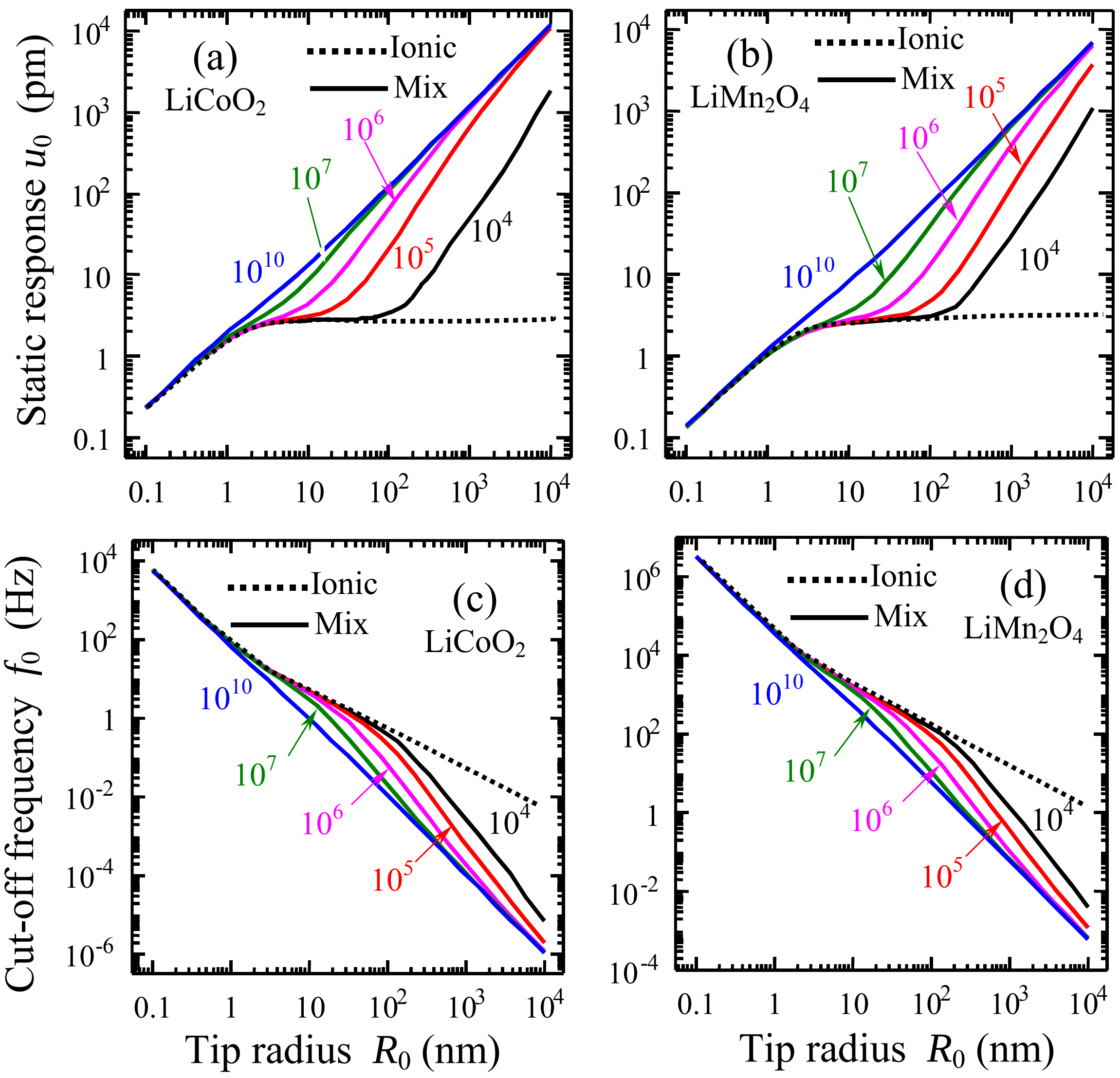

**Fig. 9.** The static response $u_0$ (a, b) and cut-off frequency $f_0 = \omega_0 / 2\pi$ (c, d) dependences on the contact radius calculated for the material parameters of $LiCoO_2$ (a, c) and $LiMnO_2$ (b, d) for different rate constant values $w/D_n = 10^4$, $10^5$, $10^6$, $10^7$, $10^{10}$ $m^{-1}$ (numbers near the curves) and fixed Debye length $R_D \approx 1$ nm. Dotted curves are purely ionic response calculated from Eqs.(9), solid curves are mixed response calculated from Eqs.(8). Material parameters of $LiCoO_2$ and $LiMn_2O_4$ are listed in the **Table 3**.

From **Fig. 9,** the static response and cut-off frequency are proportional to $R_0$ and $R_0^{-2}$ respectively and do not depend on rate constant $w$ for small $R_0 \le R_D$ (where the approximation of "purely ionic response" is applicable). At higher $R_0$ the static response and cut-off frequency show strong dependence on the rate constant, which are related to the electronic contribution. Under the condition $R_0 > R_D$ the deviation from "purely ionic response" model becomes essential, and for $R_0 >> R_D$ the following relations for static response $u_0 \approx 2(1+\nu)\beta \overline{N}_d^+ \cdot R_D (qV/k_B T)$ (independent on $R_0$) and cut-off frequency $\omega_0 \approx D_d/(R_D R_0)$ become invalid. The crossover is visible in the central part in **Fig.9,** where solid curves (mixed response) start to deviate from the dotted ones (ionic response)**.**

From **Fig. 9**, we note that the limit $w \to 0$ for any $R_0/R_D$ strongly resembles the purely ionic response, since this limit could give exactly the same $R_0$-dependences for cut-off and static response as for the purely ionic response. However, the closer look shows that corresponding expressions differ in $R_D$ definition (factor $\sqrt{2}$), also the response spectra at intermediate frequencies do depend on $D_n$ and contain additional dispersion range, which could not be described by simple purely ionic model as shown in Fig. 8.

### IV.2. Comparison of ESM with ion-blocking and ion-conductive electrodes

Two limiting cases of the ***ion-conductive*** boundary conditions were considered earlier in Ref. [43], as either fixed concentration ($\left.\delta N_d^+\right|_{z=0} = N_d V(\omega) q/(k_B T)$) or flux ($\left.\left(\partial\left(\delta N_d^+\right)/\partial z\right)\right|_{z=0} = (N_d/R_0) V(\omega) q/(k_B T)$), being proportional to the applied voltage $V(\omega)$. In the designations of the current model of ESM response for isotropic Vegard tensor, the maximal surface mechanical displacement acquires the form :

$$u_3(0,\omega) = -2(1+\nu)\beta N_d \frac{qV(\omega)}{k_B T} \begin{cases} \int_0^\infty dk \frac{R_0 J_1(kR_0)}{\left(k+\sqrt{k^2+i\omega/D_d}\right)}, & \text{fixed concentration;} \\ \int_0^\infty dk \frac{J_1(kR_0)}{\left(k+\sqrt{k^2+i\omega/D_d}\right)\sqrt{k^2+i\omega/D_d}}, & \text{fixed flux;} \end{cases} \tag{11}$$

Note, that the Debye screening length is absent in the characteristic root $\sqrt{k^2 + i\omega/D_d}$ in comparison with solution (9).

The comparison of the current ESM with ***ion-blocking electrodes*** and simplified ones from Ref. [43], where the bulk drift in the external field was neglected, but ion-conductive electrode was considered, is presented in **Fig. 10.**

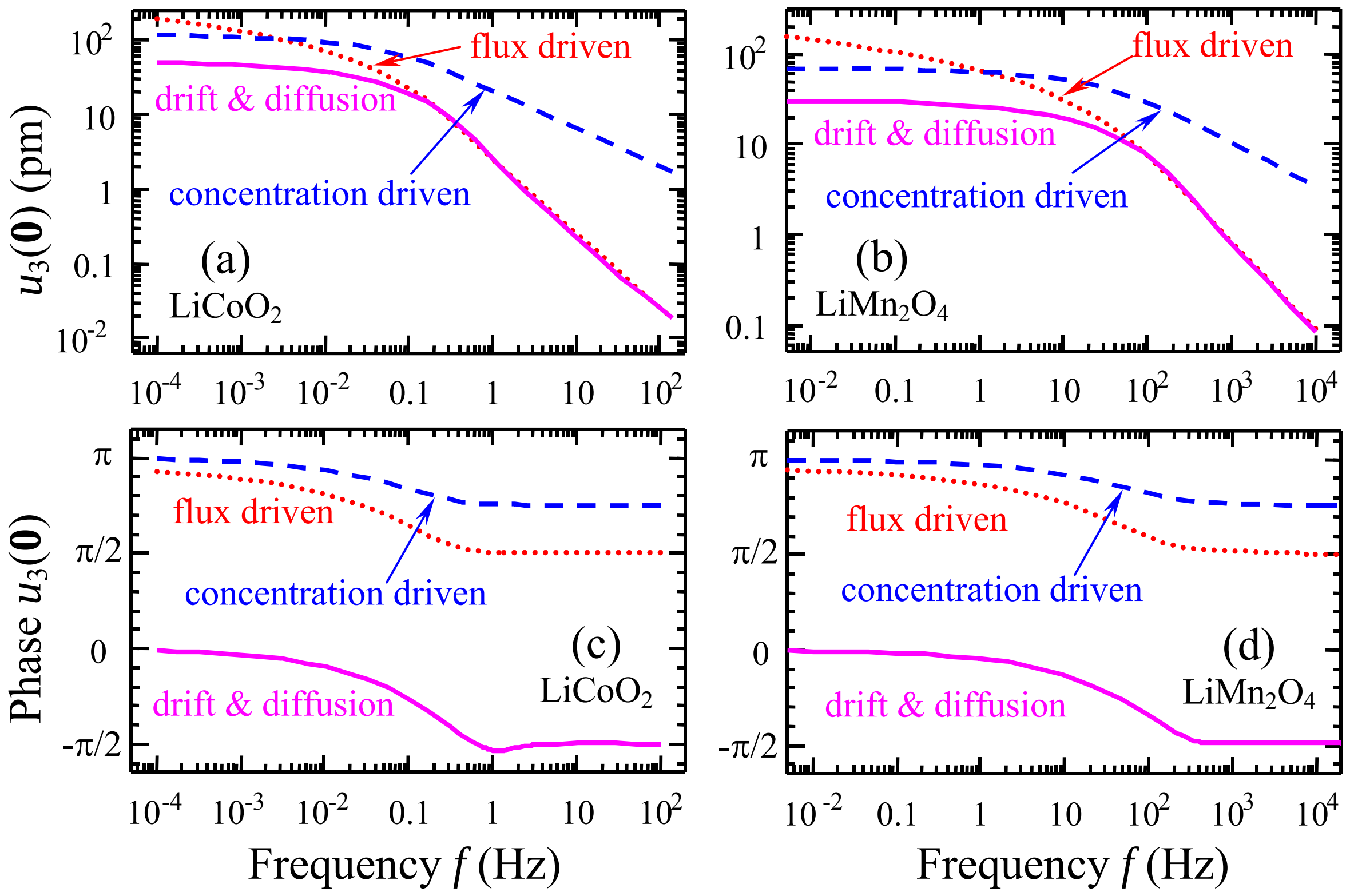

**Fig. 10.** The amplitude (a, b) and phase (c, d) of the ESM response vs. external field frequency $f$ calculated for the material parameters of $LiCoO_2$ (a, c) and $LiMn_2O_4$ (b, d) for $R_0$=50 nm and fixed Debye length $R_D \approx 1$ nm. Dotted and dashed curves are calculated on the basis of the ***ion-conductive*** ESM model from Ref.[43] (see Eqs. (10)), where only the diffusion terms are taken into account for different boundary conditions, namely for fixed value of either concentration (dashed curves) or flux (dotted curves). Solid curves are mixed ESM response with ***ion-blocking*** electrodes calculated from Eqs.(8) for diffusion coefficients ratio $D_d/D_n = 10^{-2}$ and rate constant $w/D_n = 10^8$ m$^{-1}$. Material parameters of $LiCoO_2$ and $LiMn_2O_4$ are listed in the **Table 3**.

It is seen from **Fig. 10**, that the main distinctions between the ion-blocking and ion-conducting models are phase difference about $\pi$ (i.e. signs are different); different asymptotic at small and high frequencies. In fact the solution at fixed flux diverges as $\ln(\omega)$ at small frequencies and similarly to the present model behaves as $\sim D_d / i\omega$ at the high frequencies. Concentration driven solution is finite at small frequencies but decreases as $\sqrt{D_d / i\omega}$ with frequency increase. The latter behavior resembles Warburg impedance in the case of flat capacitor geometry (see e.g. Refs. [43, 65]). Notably, for considered materials reaction-driven process yields higher responses and is preponderant at all frequencies.

## Summary

To summarize, the linear analytical theory for the calculations of the electrochemical strain microscopy response for mixed electronic-ionic conductors with blocking electrodes is developed. The solution of linearized drift and diffusion equations allows frequency dependence of response and its dependence of probe radius to be developed. The resultant behavior is compared with previously developed concentration-and flux driven ESM signals, and it is shown that the interfacial-driven process is dominant for electrochemically active probes in all frequency ranges.

Since we include both electromigration and diffusion transfer mechanisms into the calculations and obtained approximate analytical expressions for ESM response, it may seem that obtained expressions have a “universal” applicability to all materials from resistive switching memory with strongly dominant electromigration contribution to paraelectric perovskites like $SrTiO_3$ and Li-containing ionics with dominant diffusion contribution. However such conclusion is premature, since made approximations limit the quantitative applicability of the obtained expressions for all MIECs. Several limitations required for those MIECs, which local electrochemical strains can be quantitatively described quantitatively by expressions, obtained in the paper, namely:

1) Adopted linear drift-diffusion model for the ionic and electronic currents combined with assumption of constant diffusion coefficients and mobilities are conventional for Li-containing solid electrolytes and oxide perovskites, but should be used with great care for e.g. correlated oxides.

2) Analytical expressions for ESM response obtained in decoupling approximation are derived for the transversally isotropic Vegard tensor $\beta_{ij} = \delta_{ij}\beta_{ii}$ ($\beta_{11} = \beta_{22} \neq \beta_{33}$), they are not valid for anisotropic MIECs

3) We neglected the contribution of the electrostriction into the local ESM response. This is valid for semiconductors ε smaller than several tens, which electrostriction coefficients are such that the electrostriction contribution becomes essential only at high electric fields (see e.g. [[66], [67]]). Electrostriction was shown to be important for perovskites like paraelectric $SrTiO_3$ with high relative dielectric permittivity ε≥300 [[68]].

4) We consider only enough thick slabs, which local ESM response becomes virtually thickness independent. The approximation of semi-infinite MIEC used entire the paper is valid when MIEC thickness $h$ is much higher than the screening radius $R_D \sim 0.5$-50 nm, since the ESM probe electric field decay exponentially with the depth z as $\exp(-z/R_D)$.

**Acknowledgements**:

Research supported (SVK,) by the U.S. Department of Energy, Basic Energy Sciences, Materials Sciences and Engineering Division. ANM and EEA research is done by their personal costs. ANM and EEA acknowledge user share-free agreement with CNMS N UR-08-869.

## Appendix A. Equilibrium solution without applied voltage

The boundary problem for electrostatic potential distribution in the following form:

$$\begin{cases} \Delta\varphi(\mathbf{r}) = -\dfrac{q}{\varepsilon_0\varepsilon}\left(N_d^+(\mathbf{r}) - n(\mathbf{r})\right), \\ \varphi(0) = 0, \quad \varphi(h\to\infty) = 0, \quad E_z = -\left.\dfrac{d\varphi}{dz}\right|_{h\to\infty} = 0. \end{cases} \tag{A.1}$$

Static continuity equations: $\mathrm{div}J_d = -D_d\Delta N_d^+ - \eta_d\left(N_d^+\Delta\varphi + \nabla N_d^+\nabla\varphi\right) = 0$ and $\mathrm{div}J_n = +D_n\Delta n - \eta_n\left(n\Delta\varphi + \nabla n\nabla\varphi\right) = 0$ along with the boundary of the currents absence across the interfaces in the equilibrium, i.e. $J_{dz}(\rho,0,t) = -\left. D_d\dfrac{\partial}{\partial z}N_d^+ - \eta_d N_d^+\dfrac{\partial}{\partial z}\varphi\right|_{z=0} = 0$ and $J_{nz}(\rho,0) = \left. D_n\dfrac{\partial}{\partial z}n - \eta_n n\dfrac{\partial}{\partial z}\varphi\right|_{z=0} = 0$, have the only solution

$$N_d^+(\mathbf{r}) = N_0\exp\left(-\frac{q\varphi(\mathbf{r})}{k_BT}\right), \qquad n(\mathbf{r}) = n_0\exp\left(\frac{q\varphi(\mathbf{r})}{k_BT}\right) \tag{A.2}$$

Where we used the Nernst-Einstein relation $\dfrac{\eta_d}{D_d} = \dfrac{\eta_n}{D_n} = \dfrac{q}{k_BT}$. Substitution of the expressions (A.2) into the boundary problem (A.1) leads to the evident solution $\varphi(\mathbf{r}) = 0$ and $N_0 = n_0 = \overline{N}_d^+ = \overline{n}$ (local electroneutrality condition).

## Appendix B. Kinetic solution for mixed response

*B.1. Calculations of kinetic solution for the mixed response case*

Substitution $N_d^+(\mathbf{r},t) = \overline{N}_d^+ + \delta N_d^+(\mathbf{r},t)$, $n(\mathbf{r},t) = \overline{n} + \delta n(\mathbf{r},t)$ in kinetic equations leads to:

$$\frac{\partial N_d^+}{\partial t} - D_d\Delta N_d^+ - \eta_d\nabla\left(N_d^+\nabla\varphi\right) = \frac{\partial\,\delta N_d^+}{\partial t} - D_d\Delta\delta N_d^+ - \eta_d\overline{N}_d^+\Delta\varphi - \eta_d\nabla\delta N_d^+\nabla\varphi = 0, \tag{B.1a}$$

$$-\frac{\partial n}{\partial t} + D_n\Delta\delta n - \eta_n\nabla\left(n\nabla\varphi\right) = -\frac{\partial n}{\partial t} + D_n\Delta\delta n - \eta_n\overline{n}\Delta\varphi - \eta_n\nabla\delta n\nabla\varphi = 0, \tag{B.1b}$$

In the first order of perturbation theory we will neglect the terms $\nabla\delta N_d^+\nabla\varphi$ and $\nabla\delta n\nabla\varphi$ in Eqs.(B.1) as proportional to $V_0^2(\rho,\omega)\exp(2i\omega t)$. Poisson equation is

$$\Delta\varphi(\mathbf{r},t) = -\frac{q}{\varepsilon_0\varepsilon}\left(\delta N_d^+(\mathbf{r},t) - \delta n(\mathbf{r})\right). \quad \text{(B.1c)}$$

Using Fourier transformation in time and x,y-domain in Eqs.(B.1) we get:

$$\left(i\omega + k^2 D_d\right)\tilde{N}_d^+ - D_d \frac{\partial^2}{\partial z^2}\tilde{N}_d^+ + \eta_d \bar{N}_d^+ \frac{q}{\varepsilon_0\varepsilon}\left(\tilde{N}_d^+ - \tilde{n}\right) = 0, \quad \text{(B.2a)}$$

$$-\left(i\omega + k^2 D_n\right)\tilde{n} + D_n \frac{\partial^2}{\partial z^2}\tilde{n} + \eta_n \bar{n}\frac{q}{\varepsilon_0\varepsilon}\left(\tilde{N}_d^+ - \tilde{n}\right) = 0, \quad \text{(B.2b)}$$

$$\left(\frac{\partial^2}{\partial z^2} - k^2\right)\tilde{\varphi}(\mathbf{r}) = -\frac{q}{\varepsilon_0\varepsilon}\left(\tilde{N}_d^+ - \tilde{n}\right) \quad \text{(B.2c)}$$

The solution of the system (B.2) allowing for the boundary conditions (4) as:

$$\tilde{N}_d^+(k,z,\omega) = \sum_{i=1,2} N_i(k,\omega)\exp(-s_i(k,\omega)z), \quad \text{(B.3a)}$$

$$\tilde{n}(k,z,\omega) = \sum_{i=1,2}\left(1 - \tau_d\left(D_d\left(s_i^2(k,\omega) - k^2\right) - i\omega\right)\right)N_i(k,\omega)\exp(-s_i(k,\omega)z), \quad \text{(B.3b)}$$

$$\tilde{\varphi}(k,z,\omega) = \psi(k,\omega)\exp(-kz) + \frac{q\tau_d}{\varepsilon_0\varepsilon}\sum_{i=1,2}\left(\frac{i\omega}{s_i^2 - k^2} - D_d\right)N_i(k,\omega)\exp(-s_i(k,\omega)z). \quad \text{(B.3c)}$$

eigen values

$$s_{1,2}^2(k,\omega) = k^2 + \frac{1}{2}\left(\left(i\omega + \frac{1}{\tau_d}\right)\frac{1}{D_d} + \left(i\omega + \frac{1}{\tau_n}\right)\frac{1}{D_n}\right) \pm$$
$$\pm\frac{1}{2}\sqrt{\left(\left(i\omega + \frac{1}{\tau_d}\right)\frac{1}{D_d} + \left(i\omega + \frac{1}{\tau_n}\right)\frac{1}{D_n}\right)^2 - \frac{4i\omega}{D_d D_n}\left(i\omega + \frac{1}{\tau_d} + \frac{1}{\tau_n}\right)}, \quad \text{(B.4)}$$

Maxwellian times are $\tau_d = \frac{\varepsilon_0\varepsilon}{\eta_d \bar{N}_d^+ q}$ and $\tau_n = \frac{\varepsilon_0\varepsilon}{\eta_n \bar{n} q}$. Under the validity of the Planck-Nernst-Einstein relation $\frac{\eta_d}{D_d} = \frac{\eta_n}{D_n} = \frac{q}{k_B T}$ and the electroneutrality condition $\bar{N}_d^+ = \bar{n}$ for the system in the equilibrium, the Debye screening radius acquires can be introduced as $R_D = \sqrt{\tau_d D_d} = \sqrt{\tau_n D_n} = \sqrt{\frac{\varepsilon_0\varepsilon D_d}{\eta_d \bar{N}_d^+ q}} = \sqrt{\frac{\varepsilon_0\varepsilon D_n}{\eta_n \bar{n} q}}$. The Eq.(B.4) acquires the form:

$$s_{1,2}(k,\omega) = \sqrt{k^2 + \frac{i\omega}{2}\left(\frac{1}{D_n} + \frac{1}{D_d}\right) + \frac{1}{R_D^2} \pm \sqrt{\frac{1}{R_D^4} - \frac{\omega^2}{4}\left(\frac{1}{D_d} - \frac{1}{D_n}\right)^2}}.$$

The amplitudes have form:

$$N_i(k,\omega)=\varepsilon_0\varepsilon\widetilde{V}_0(k)\frac{a_i(k,\omega)-wb_i(k,\omega)}{A(k,\omega)-wB(k,\omega)}, \quad \text{(B.5a)}$$

$$\psi(k,\omega)=i\omega\widetilde{V}_0(k)\frac{c(k,\omega)-wd(k,\omega)}{A(k,\omega)-wB(k,\omega)}\frac{\tau_d q(s_2-s_1)}{(k-s_1)(k-s_2)}. \quad \text{(B.5b)}$$

Where

$$A=q(s_2-s_1)\tau_d\begin{pmatrix}D_d k(k+s_1)(k+s_2)(D_n\tau_n+D_d\tau_d(1+D_n(k^2-s_1^2-s_1s_2-s_2^2)\tau_n))+\\ i(k+s_1+s_2)(D_n\tau_n+D_d\tau_d(1+D_n(2k^2-s_1^2-s_2^2)\tau_n))\omega-D_n(k+s_1+s_2)\tau_d\tau_n\omega^2\end{pmatrix},$$

$$B=q(s_1-s_2)\tau_d\tau_n\begin{pmatrix}i(1+D_d(k^2-s_1^2-s_1s_2-s_2^2-k(s_1+s_2))\tau_d)\omega\\ -D_d^2k(k+s_1)(k+s_2)(s_1+s_2)\tau_d-\tau_d\omega^2\end{pmatrix}.$$

(B.6a)

$$\begin{aligned}
a_1(k,\omega)&=-ks_2(k+s_1)(k+s_2)(D_n\tau_n+\tau_d(D_d+D_dD_n\tau_n(k^2-s_2^2)+iD_n\tau_n\omega)),\\
b_1(k,\omega)&=ks_2(k+s_1)(k+s_2)D_n\tau_n(1+D_d\tau_d(k^2-s_2^2)+i\tau_d\omega),\\
a_2(k,\omega)&=ks_1(k+s_1)(k+s_2)(D_n\tau_n+\tau_d(D_d+D_dD_n\tau_n(k^2-s_1^2)+iD_n\tau_n\omega)),\\
b_2(k,\omega)&=-ks_1(k+s_1)(k+s_2)D_n\tau_n(1+D_d\tau_d(k^2-s_1^2)+i\tau_d\omega)
\end{aligned} \quad \text{(B.6b)}$$

$$\begin{aligned}
c(k,\omega)&=s_1s_2(s_1+s_2)(D_d\tau_d+D_n\tau_n+iD_n\tau_n\omega\tau_d+D_d\tau_dD_n\tau_n(2k^2-s_1^2-s_2^2)),\\
d(k,\omega)&=-D_n\tau_n\left(D_d\tau_d\left((k^2+s_1s_2)^2-s_1s_2(s_1+s_2)^2\right)+(k^2+s_1s_2)(1+i\tau_d\omega)\right)
\end{aligned} \quad \text{(B.6c)}$$

*__B.2. Calculations of kinetic solution of the purely ionic case__*

The purely ionic response of mixed electronic-ionic conductors corresponds to the physical case, when there are a lot of electrons and they are free to move. Probably, $LiCoO_2$ in the metallic phase qualify here.

Under the condition $D_n >> D_d$ one could neglect electron charge variations in Eqs.(B.2).

Then, using Fourier transformation in time and *x*,*y*-domain in Eqs.(B.2) we get

$$(i\omega+k^2D_d)\delta\widetilde{N}_d^+(k,z,\omega)-D_d\frac{\partial^2}{\partial z^2}\delta\widetilde{N}_d^+(k,z,\omega)+\eta_d\overline{N}_d^+\frac{q}{\varepsilon_0\varepsilon}\delta\widetilde{N}_d^+(k,z,\omega)=0, \quad \text{(B.7a)}$$

$$\left(\frac{\partial^2}{\partial z^2}-k^2\right)\widetilde{\varphi}(k,z,\omega)=-\frac{q}{\varepsilon_0\varepsilon}\delta\widetilde{N}_d^+(k,z,\omega). \quad \text{(B.7b)}$$

Boundary conditions

$$\widetilde{\varphi}(k,0,\omega)=\widetilde{V}_0(k,\omega),\quad \widetilde{\varphi}(k,h\to\infty,\omega)=0\,, \tag{B.7c}$$

$$\begin{aligned}
&-D_d\frac{\partial}{\partial z}\delta\widetilde{N}_d^+(k,0,\omega)-\eta_d\overline{N}_d^+\frac{\partial}{\partial z}\widetilde{\varphi}(k,0,\omega)=0,\\
&-D_d\frac{\partial}{\partial z}\delta\widetilde{N}_d^+(k,h,\omega)-\eta_d\overline{N}_d^+\frac{\partial}{\partial z}\widetilde{\varphi}(k,h,\omega)\bigg|_{h\to\infty}=0.
\end{aligned} \tag{B.7d}$$

In the high thickness limit $h\to\infty$ the solution of Eqs.(B.7) in the form of 2D-Fourier images can be simplified as:

$$\delta\widetilde{N}_d^+(k,z,\omega)=-\overline{N}_d^+\widetilde{V}_0(k,\omega)\frac{\eta_d}{D_d}\frac{k\left(1+i\omega\left(R_D^2/D_d\right)\right)\exp(-s(k,\omega)z)}{k+i\omega\left(R_D^2/D_d\right)s(k,\omega)}\,, \tag{B.8a}$$

$$\widetilde{\varphi}(k,z,\omega)=\widetilde{V}_0(k,\omega)\left(\frac{k\exp(-s(k,\omega)z)}{k+i\omega\left(R_D^2/D_d\right)s(k,\omega)}+\frac{i\omega\left(R_D^2/D_d\right)s(k,\omega)\exp(-kz)}{k+i\omega\left(R_D^2/D_d\right)s(k,\omega)}\right). \tag{B.8b}$$

Note that as per kinetic equation (B.7a) and boundary condition (8B.b) the total quantity of donors is independent on time, $\partial\int_V N_d^+dV\big/\partial t=0$, which means that $i\omega\int_V\delta\widetilde{N}_d^+(k=0,z,\omega)dz=0$.

Where the characteristic eigenvalue $s(k,\omega)$ is introduced as

$$s(k,\omega)=\sqrt{k^2+\frac{i\omega}{D_d}+\frac{1}{R_D^2}}\,. \tag{B.9}$$

## Appendix C. ESM response calculations

$$\begin{aligned}
u_3(\rho,0,\omega)&=-\int_{-\infty}^{\infty}dk_x\int_{-\infty}^{\infty}dk_y\int_0^{\infty}dz e^{-ik_xx-ik_yx-kz}\delta N_d^+(k,z,\omega)\left(\frac{\beta_{33}}{2\pi}(1+kz)+\frac{\beta_{11}}{2\pi}(1+2\nu-kz)\right)\\
&=-\int_0^{\infty}dkJ_0(k\rho)k\int_0^{\infty}dz(\beta_{33}(1+kz)+\beta_{11}(1+2\nu-kz))\sum_{m=1,2}N_m(k,\omega)\exp(-s_m(k,\omega)z-kz)\\
&=-\int_0^{\infty}dkJ_0(k\rho)k\sum_{m=1,2}N_m(k,\omega)\left(\beta_{33}\frac{2k+s_m}{(k+s_m)^2}+\beta_{11}\frac{2k\nu+s_m(1+2\nu)}{(k+s_m)^2}\right)
\end{aligned}$$

(C.1)

Substitution of Eq.(B.8a) into Eq.(C.1) for the case $\beta_{ii}=\beta$ leads to the expression

$$u_3(\rho,\omega) = 2(1+\nu)\beta\overline{N}_d^+ \frac{\eta_d}{D_d}\int_0^\infty dk \frac{k^2\left(1+i\omega\left(R_D^2/D_d\right)\right)\widetilde{V}_0(k,\omega)J_0(k\rho)}{\left(k+s(k,\omega)\right)\left(k+i\omega\left(R_D^2/D_d\right)s(k,\omega)\right)}, \tag{C.2}$$

Since $2(1+\nu)\beta\overline{N}_d^+ \frac{qV(\omega)}{k_BT}\int_0^\infty dk \frac{R_0 J_1(kR_0)J_0(kR_0)}{k+\sqrt{k^2+\frac{1}{R_D^2}}} \approx 2(1+\nu)\beta\overline{N}_d^+ \frac{qV(\omega)}{k_BT}\left(\frac{R_0R_D}{2R_0+\pi R_D}\right)$, response distribution is

$$u_3(\rho,\omega) = 2(1+\nu)\beta\overline{N}_d^+ \frac{\eta_d}{D_d}V(\omega)\int_0^\infty dk \frac{kR_0J_1(kR_0)J_0(k\rho)}{\left(k\left(k+\sqrt{k^2+\frac{i\omega}{D_d}+\frac{1}{R_D^2}}\right)+\frac{i\omega}{D_d}\right)} \tag{C.3a}$$

Maximal response is

$$u_3(\rho=0,\omega) = 2(1+\nu)\beta\overline{N}_d^+ \frac{\eta_d}{D_d}V(\omega)\int_0^\infty dk \frac{kR_0J_1(kR_0)}{\left(k\left(k+\sqrt{k^2+\frac{i\omega}{D_d}+\frac{1}{R_D^2}}\right)+\frac{i\omega}{D_d}\right)} \tag{C.3b}$$

Response averaged on the tip area is

$$\left\langle u_3(\rho,\omega)\right\rangle_\rho = 2(1+\nu)\beta\overline{N}_d^+ \frac{\eta_d}{D_d}V(\omega)\int_0^\infty dk \frac{2J_1(kR_0)^2}{\left(k\left(k+\sqrt{k^2+\frac{i\omega}{D_d}+\frac{1}{R_D^2}}\right)+\frac{i\omega}{D_d}\right)} \tag{C.3c}$$

Response at the contact line is

$$u_3(\rho=R_0,\omega) = 2(1+\nu)\beta\overline{N}_d^+ \frac{\eta_d}{D_d}V(\omega)\int_0^\infty dk \frac{kR_0J_1(kR_0)J_0(kR_0)}{\left(k\left(k+\sqrt{k^2+\frac{i\omega}{D_d}+\frac{1}{R_D^2}}\right)+\frac{i\omega}{D_d}\right)} =$$

$$\left|\begin{matrix} R_0 dk\, J_1(kR_0)J_0(kR_0) = \\ d\left(-J_0(kR_0)^2\right)/2 \end{matrix}\right| = \left\{\text{integration on parts}\right\} = \tag{C.3d}$$

$$= 2(1+\nu)\beta\overline{N}_d^+ \frac{\eta_d}{D_d}V(\omega)\int_0^\infty dk \frac{\frac{J_0(kR_0)^2}{2}\left(\left(-k^2+\frac{i\omega}{D_d}\right)\sqrt{k^2+\frac{i\omega}{D_d}+\frac{1}{R_D^2}}-k^3\right)}{\sqrt{k^2+\frac{i\omega}{D_d}+\frac{1}{R_D^2}}\left(k\left(k+\sqrt{k^2+\frac{i\omega}{D_d}+\frac{1}{R_D^2}}\right)+\frac{i\omega}{D_d}\right)^2}$$

*Remark to Fig.8.* For purely ionic response $u_0 = 2(1+\nu)\beta V \frac{R_D R_0}{(2R_D + R_0)} \frac{q\overline{N}_d^+}{k_B T}$ and $\omega_0 = \frac{2(1+\nu)\beta V}{u_0} \frac{D_d}{R_0} \frac{q\overline{N}_d^+}{k_B T} = \frac{(2R_D + R_0)}{R_D} \frac{D_d}{R_0^2}$